\documentclass[11pt,a4paper]{article}
\usepackage{epsfig}
\usepackage{graphics}
\usepackage{subfigure}
\usepackage{amssymb}
\usepackage{amsmath}
\usepackage{amsfonts}

\makeatletter

\@addtoreset{equation}{section}
\makeatother

\def\vecs{{\pmb{\sigma}}}

\def\vectau{{\pmb{\tau}}}

\def\vecpi{{\pmb{\pi}}}

\begin{document}
\title{
  \vskip 15pt
\rightline{{\small Leeds Pure Mathematics 2002/24}}
\vspace{1cm} 
  {\bf \Large \bf Fermions coupled to Skyrmions on $S^3$}
  \vskip 10pt}
\author{ 
Steffen Krusch\thanks{S.Krusch@maths.leeds.ac.uk } \\[5pt]
{\normalsize {\sl Department of Pure Mathematics}}\\
{\normalsize {\sl School of Mathematics, University of Leeds, Leeds LS2 
9JT}}\\ 
{\normalsize {\sl England}}
}

\date{April 30, 2003}
 
\maketitle

\vspace{2cm}

\begin{abstract}
\noindent
This paper discusses Skyrmions on the $3$-sphere coupled to
fermions. The resulting Dirac equation commutes with a generalized angular
momentum ${\bf G}$. For $G = 0$ the Dirac equation can be solved
explicitly for a constant Skyrme configuration and also for a
$SO(4)$ symmetric hedgehog configuration. 
We discuss how the spectrum changes due to the presence of a non-trivial 
winding number, and also consider more general Skyrme configurations 
numerically.
\end{abstract}

\vspace{5cm}
\begin{tt}
\pageref{lastref} pages, 6 figures
\end{tt}

\newpage
\section{Introduction}

The Skyrme model is a classical field theory which describes
atomic nuclei and their low
energy interactions \cite{Skyrme:1961vq}. 
It is a non-linear theory admitting
topological solitons, 
so-called Skyrmions, labelled by a topological charge $B$ which can be 
identified with the baryon number.  
When the theory is quantized the Skyrmions give a
good description of nuclei, $\Delta$ resonance \cite{Adkins:1983ya} and 
even bound states of nuclei \cite{Leese:1995hb, Irwin:1998bs,
Krusch:2002by}. 
A striking feature is that Skyrmions can be quantized as fermions 
\cite{Finkelstein:1968hy,Witten:1983tw}.  
When the theory is coupled to
a fermion field two different ways of describing fermions appear in
the same theory. Physically, the fermion field can be thought of as light
quarks in the presence of atomic nuclei \cite{Balachandran:1998zq}. 
Hiller calculated the spectrum of
the fermions in the background of a Skyrmion \cite{Hiller:1986ry}. 
However, the 
approximation for the Skyrmion configuration was very crude. The
spectrum shows a spectral flow of the eigenvalues depending on the
value of the coupling constant. In particular one mode crosses
zero. This general behaviour has been verified by other authors
\cite{Kahana:1984be,Ripka:1985am, Ripka:1997}. 
To gain a better understanding we consider a generalization of the
model.
Since static field configurations in the original Skyrme model 
in flat space are topologically
equivalent to maps $S^3 \to S^3$, the model can be  generalized to the
base space being a sphere of radius $L$. 
The original model is recovered in the limit
$L \to \infty$. It turns out that for $L=1$ and baryon number $B=1$
the Faddeev-Bogomolny bound is satisfied and the minimal energy solution can
be worked out explicitly \cite{Manton:1986pz}. 
In this case the map $S^3 \to S^3$ is an isometry. 
For larger $L$ the Skyrmions are localized and tend to the usual 
flat space Skyrmions in the limit $L \to \infty$. 

Monopoles and Skyrmions have many properties in common, and it is
therefore instructive to recall the results for monopoles. There is an
index theorem making 
use of the Bogomolny equations which states that for $n$ monopoles
there is a $n$ dimensional vector space of fermionic zero modes
\cite{Manton:1993aa}. 
Furthermore, the monopole and its fermionic zero mode can be
calculated explicitly for $n=1$. We will examine how
far we can extend the analogy between monopoles and Skyrmions.

In this paper we study fermions coupled to Skyrmions on $S^3$. 
The Dirac equation on $S^3$ is derived in Sect. 
\ref{DiracEquation}. We also explain the Skyrmion-fermion interaction. 
In Sect. \ref{DiracSolution}, 
we derive the ordinary differential equation which
describes the solutions with total angular momentum $G = 0$
for general shape functions. In Sect. \ref{DifferentialEquations}, we
give a brief discussion of ordinary differential equations and their regular
singular points.
In Sect. \ref{ConstantSkyrmion}, the equation of Sect. 
\ref{DiracSolution} is solved for a constant Skyrme configuration, 
which is equivalent to a mass term for the fermions. 
We also discuss the boundary conditions which are
necessary to derive the spectrum. 
In Sect. \ref{HedgehogSkyrmion}, we derive the eigenfunctions and
the spectrum for the hedgehog configuration with the special shape
function $f(\mu) = \mu$. 
In Sect. \ref{Numerics} the results of Sect.
\ref{ConstantSkyrmion} and
\ref{HedgehogSkyrmion}  are verified numerically. 
The spectrum for more general shape functions is also evaluated 
numerically and compared with results in the literature. 
We find an interesting relationship between the topological charge of 
the Skyrmion and certain singular points. We end with a conclusion.

\section{The Dirac Equation on $S^3$}
\label{DiracEquation}

In this section we derive the Dirac equation on a $3$-sphere of radius
$L=1$. 
First we define an atlas of two charts $X_i$ and ${\tilde X}_i$ which
are the stereographic projections from the north pole $N$ and the
south pole $S$, respectively. 
Next we define vierbeins on both charts which uniquely determine a
connection one-form using the metric compatibility condition and the
torsion-free condition. The connection one-form induces the spin
connection so that we can finally write down the Dirac equation for
each chart.\footnote{A good discussion of frame fields and
spin-connections can be found in \cite{Nakahara:1990}, 
and also in \cite{Eguchi:1980jx}.} 
At the end of this section we discuss the interaction term which
couples fermions and Skyrmions.

In order to construct the stereographic projection let $S^3$ 
be embedded in Euclidean ${\mathbb R}^4$ with coordinates
$(x_1,x_2,x_3,w)$. 
The points of $S^3$ can now be labelled by the
projection from the north pole $N$ onto the ${\mathbb R}^3$ plane through
the equator introducing the coordinates $X_i$. Alternatively, one
can project from the south pole $S$ introducing the coordinates
${\tilde X}_i$. 
These two charts are well defined everywhere apart from their
projection point $N$ and $S$, respectively.
In terms of the ${\mathbb R}^4$ Cartesian coordinates we obtain
\begin{equation}
X_i = \frac{x_i}{1-w} 
{\rm ~~~and~~~}
{\tilde X}_i = \frac{x_i}{1+w}.
\end{equation}

In the following, we derive the equations of motion for both charts. 
As we shall see in Sect. \ref{ConstantSkyrmion}, this enables us to discuss the 
regularity conditions at the north and south pole in a simple way. 
We will also use the two charts to discuss a nontrivial symmetry of the 
energy spectrum in Sect. \ref{Generalization}.

The induced metric on $S^3$
can be calculated from the Euclidean metric in ${\mathbb R}^4$.
Taking into account that our base space is a
Lorentzian space time ${\mathbb R} \times S^3$, the metric can be written
as 
\begin{equation}
g = {\rm diag} \left(1, \frac{-4}{1+R^2},\frac{-4}{1+R^2},\frac{-4}{1+R^2}
\right),\ 
{\rm where}\
R^2 = X_1^2+X_2^2+X_3^2.
\end{equation}
The other coordinate system gives rise to the same metric replacing
$X_i$ by ${\tilde X}_i$. On the overlap the two charts are
related via the transition functions
\begin{equation}
{\tilde X}_i = \frac{1}{R^2} X_i.
\end{equation}
Therefore, the coordinate vectors $\partial_{X_i}$ and $\partial_{{\tilde X}_i}$ 
are related by
\begin{equation} 
\partial_{{\tilde X}_i} = \left(R^2 \delta_{ij} - 2 X_i X_j \right)
\partial_{X_j}.
\end{equation}
Note that this transformation has the determinant ${\rm det}
(\frac{\partial X_i}{\partial {\tilde X}_j}) = -R^6$.
We now choose non-coordinate bases ${\hat e}_\alpha = {e_\alpha}^\mu
\partial_{X_\mu}$ and ${\hat {\tilde e}}_\alpha = {{{\tilde
e}_\alpha}}^{~\mu} \partial_{{\tilde X}_\mu}$. 
It is convenient to choose diagonal vierbeins
${e_\alpha}^\mu$ and ${{\tilde e}_\alpha}^{~\mu}$ such that
\begin{eqnarray}
\begin{array}{c}
{e_0}^0 = 1, \\ \\
{e_i}^i = - \frac{1+R^2}{2},
\end{array}
{\rm ~~~and~~~}
\begin{array}{c}
{{\tilde e}_0}^{~0} = 1, \\ \\
{{\tilde e}_i}^{~i} =  \frac{1+{\tilde R}^2}{2},
\end{array}
\end{eqnarray}
where all other components vanish. The minus sign occurs because
$X_i$ and ${\tilde X}_i$ are related by inversion which changes the
orientation. The vierbeins satisfy
\begin{equation}
  {e_\alpha}^\mu {e_\beta}^\nu g_{\mu \nu} = \eta_{\alpha \beta},
\end{equation}
where $\eta_{\alpha \beta} = {\rm diag}(1,-1,-1,-1)$ is the flat Minkowski
metric. The same relations are true for the vierbeins ${{\tilde e}_\alpha}^{~\mu}$.
The non-coordinate basis vectors are related by an $SO(3)$-transformation
\begin{equation}
\label{erotation}
{\hat {\tilde e}}_i = \frac{1}{R^2} \left( 2 X_i X_j - \delta_{i j}\right) {\hat
e}_j.
\end{equation}
This is in fact a $\pi$-rotation about the vector $\frac{X_i}{R}$.  
With this choice of vierbeins the connection 
one-form $\omega_{\alpha
\beta}$ can be calculated from the metric compatibility condition 
$\omega_{\alpha \beta} = - \omega_{\beta \alpha}$ and the torsion-free
condition 
\begin{equation}
{\rm d} {\hat \theta}^\alpha + {\omega^\alpha}_\beta \wedge {\hat
\theta}^\beta = 0,
\end{equation}
where ${\hat \theta}^\alpha = {e^\alpha}_\mu {\rm d} X^\mu$ is the
dual basis of ${\hat e}_\alpha$. 
After a short calculation we obtain
\begin{equation}
\omega^{\alpha \beta} = \frac{2}{1+R^2} \left( X^\alpha {\rm d}
X^\beta - X^\beta {\rm d} X^\alpha \right),
\end{equation}
and the same relation holds for ${\omega}^{\alpha \beta}$ expressed in
${\tilde X}_i$-coordinates.
The spin connection $\Omega_\mu$ is now given by
\begin{equation}
   \Omega_\mu {\rm d} X^\mu = -\tfrac{i}{2}\, \omega^{\alpha \beta}
   \Sigma_{\alpha \beta},
\end{equation}
where $\Sigma_{\alpha \beta} = \frac{i}{4} \left[ \gamma_\alpha,
\gamma_\beta \right]$, and the $\gamma$-matrices satisfy the standard
anticommutation relations 
$\{\gamma_\alpha,\gamma_\beta \} = 2 \eta_{\alpha \beta}$.
The Lagrangian for fermions in curved space time is given by 
\begin{equation}
{\cal L}_{{\rm fermion}} = {\overline \psi} \left( i \gamma^\alpha 
{e_\alpha}^\kappa \left(
\partial_\kappa + \Omega_\kappa \right) 
\right) \psi.
\end{equation}
With our choice of coordinates and vierbeins we obtain
\begin{equation}
\label{L1}
{\cal L}_{{\rm fermion}} = {\overline \psi} \left(X_i,t \right) 
\left( i \gamma^0 \partial_t - 
i \gamma^i \left(\frac{1+R^2}{2} \partial_{X_i} - X_i \right)
\right)
\psi \left(X_i,t \right),
\end{equation}
using the coordinates $X_i$ and
\begin{equation}
\label{L2}
{\cal L}_{{\rm fermion}} = {\overline {\tilde \psi}} \big({\tilde X}_i,t \big) 
\left( i \gamma^0 \partial_t + 
i \gamma^i \left(\frac{1+{\tilde R}^2}{2} \partial_{{\tilde X}_i} - 
{\tilde X}_i \right)
\right)
{\tilde \psi} \big({\tilde X}_i,t \big),
\end{equation}
for the coordinates ${\tilde X}_i$.
Equation (\ref{L2}) can be transformed into (\ref{L1}) by
using ${\tilde X}_i = \frac{1}{R^2} X_i$ and the induced spinor
transformation $\rho$  of the fermion field $\psi(X_i,t)$ which we
calculate in the following. The spinor transformation $\rho$ is defined as
\begin{equation}
{\tilde \psi}\big({\tilde X}_i\big) = \rho\, \psi \left( X_i \right).
\end{equation}
The transformation $\rho$ is a lifting of the local Lorentz
transformation $T$ of the vierbeins in (\ref{erotation}). 
Note that the transformation $T_{ij} = (2 {\hat X}_i {\hat X}_j - \delta_{i
j})$ can be rewritten as 
\begin{equation}
T(\alpha)_{ij} = \cos \alpha~ \delta_{i j} + \sin
\alpha~\epsilon_{ijk} {\hat X}_k + {\hat X}_i {\hat X}_j (1 - \cos \alpha)
\end{equation}
for $\alpha =  \pi$. Here ${\hat X}_j = \frac{X_j}{R}$.
Since we are only concerned with the spatial part we work with
the Euclidean $\delta_{ij}$ rather than with $\eta_{ij}$.
Infinitesimally, the rotation is given by $T(\alpha)_{ij} = \delta_{i j} + 
\alpha \epsilon_{i j k} {\hat X}_k$. 
This can be lifted to a spinor transformation by
$\rho(\alpha) = \exp( -\frac{i}{2} \alpha \epsilon_{i j k} {\hat X}_k
\Sigma_{ij})$. For $\alpha = \pi$ we obtain
\begin{equation}
\label{rhospin}
\rho  = i \gamma_0 \gamma_5 \gamma_j {\hat X}_j.
\end{equation}
Note that there is a sign choice. As the spinor transformation
$\rho$ is an element of the double cover of the rotation group, 
the transformation $-\rho$ is also a lifting of the rotation 
$T_{i j}$. 

In the following we discuss fermions coupled to Skyrmions. The
full Lagrangian ${\cal L}$ of this model is the sum of the fermion 
Lagrangian ${\cal L}_{{\rm fermion}}$, the Skyrmion Lagrangian ${\cal 
L}_{{\rm Skyrmion}}$ and the interaction Lagrangian ${\cal L}_{{\rm int.}}$. In 
this paper we consider fermions in the background of a static Skyrme 
configuration and neglect the backreaction. Therefore, we do not discuss 
the Skyrmion Lagrangian any further, and the interested reader is referred 
to \cite{Krusch:2000gb}. The interaction Lagrangian ${\cal L}_{{\rm int.}}$ 
given by 
\begin{eqnarray}
\label{Lint}
{\cal L}_{{\rm int.}} = - g {\overline \psi}  \left(
\sigma + i \gamma_5 \vectau \cdot \vecpi
\right) \psi,
\end{eqnarray}
where $U = \sigma + i \vectau \cdot \vecpi$ is a convenient 
parameterization of the Skyrme field.
Equation (\ref{Lint}) is the standard parity-invariant manner in which 
fermions couple to a Skyrme field and is also called the Yukawa 
interaction term. 
This Lagrangian was first introduced by
Gell-Mann and Levy (without a fourth order Skyrme term),
\cite{Gell-Mann:1960np}. Also see \cite{Ripka:1997} for a discussion
of various different chiral models. Recently, the Lagrangian
(\ref{Lint}) has been discussed by Balachandran {\it et al.}
\cite{Balachandran:1998zq}. 

Note that $\psi$ is a spin and an isospin spinor. In our notation the Pauli
matrices $\vectau$ act on the isospin indices, whereas the
$\gamma$-matrices act on the spin indices. 
It is convenient to split the Dirac spinor into two
$2 \times 2$ spin isospin matrices $\psi_{1}$ and $\psi_{2}$ such that
\begin{eqnarray}
\psi = 
\left(
\begin{array}{c}
\psi_1 \\
\psi_2
\end{array}
\right).
\end{eqnarray}
In this notation the $\gamma$-matrices can be written 
in terms of quaternionic $2 \times 2$ matrices using $i$ times the spin
Pauli matrices $\vecs$ and the two dimensional identity matrix $I_2$
as a basis for the quaternions. 
Since any complex $2 \times 2$ matrix can also be expressed 
in terms of quaternions with complex coefficients 
it is convenient to choose the same basis
for the quaternions as before. Then the spin isospin
matrices $\psi_i$ can be written as $\psi_i = a_0^{(i)} I_2 + i a_k^{(i)}
\sigma_k$. With this notation the spin operators act on $\psi$ by
left-multiplication, $\sigma_k \psi_i$, whereas the isospin matrices
act on $\psi$ by right multiplication 
\begin{eqnarray}
\label{quat1}
\tau_k \psi_i &=& \psi_i \sigma_k^T, \\
\label{quat2}
 &=& - \psi_i \sigma_2 \sigma_k \sigma_2,
\end{eqnarray} 
where the transpose in (\ref{quat1}) guarantees the group structure,
and (\ref{quat2}) gives a useful trick to evaluate $\sigma_i^T$.

As mentioned above, we make the approximation that the Skyrmion is a fixed
static background field, {\it i.e.} we ignore the backreaction of the
fermion fields.
Furthermore, we only consider spherically symmetric
Skyrmions. For $B=1$, Skyrmions are believed to be spherically symmetric, 
but for $B>1$ this is no longer true. However, we expect that some generic 
features of the Skyrmion-fermion interaction are not sensitive to the 
precise shape of the Skyrmion. For more details see Sect. 
\ref{Generalization}. 
Spherically symmetric Skyrme fields are best written in polar coordinates
\begin{equation}
U = \exp \left( i f(\mu)~ {\bf e}_\mu \cdot \vectau \right),
\end{equation}
where $f(\mu)$ is the ``radial'' shape function and ${\bf e}_\mu$ is
the normal vector in $\mu$-direction, see (\ref{unitvectors}). 
The shape function $f(\mu)$ can be
calculated by solving the Euler-Lagrange equation of the Skyrme
Lagrangian numerically. 
However, instead of using a numerical solution it is more convenient
in this context to approximate $f(\mu)$ by an explicit function.  
Manton showed in \cite{Manton:1987xt} that a good ansatz for $L$ not too large 
is the conformal ansatz given by\footnote{For 
$L \to \infty$ the ansatz tends to $f(r) = 2 \arctan \frac{r}{r_0}$. 
Qualitatively, this is still a good ansatz.} 
\begin{equation}
\label{conformal}
f(\mu) = 2 \arctan \left( k \tan \frac{\mu}{2} \right).
\end{equation}
Here $k$ is a constant which has to be determined by minimizing the
energy of the Skyrmion. 
This ansatz describes a Skyrme configuration which is either localized around
one of the poles or uniformly distributed. The value 
$k=1$ corresponds to the $O(4)$-symmetric ansatz of the previous section, 
whereas for $k \ll 1$, or $k \gg 1$, the Skyrme field is localized at the
south pole, or the north pole, respectively. 
Manton also showed that for $L \leq \sqrt{2}$ the stable solution is $k=1$.  

Finally, we can write down the Dirac equation for fermions
coupled to a spherically symmetric background Skyrmion. It is given by
\begin{equation}
\label{Dirac1}
\left( i \gamma^0 \partial_t - 
i \gamma^i \left(\frac{1+R^2}{2} \partial_{X_i} - X_i \right)
- g U^{\gamma_5}
\right)
\psi \big(X_i,t \big) = 0, 
\end{equation}
using the coordinates $X_i$ and
\begin{equation}
\label{Dirac2}
\left( i \gamma^0 \partial_t + 
i \gamma^i \left(\frac{1+{\tilde R}^2}{2} \partial_{{\tilde X}_i} - 
{\tilde X}_i \right)
-g U^{\gamma_5}
\right)
{\tilde \psi} \big( {\tilde X}_i,t \big) = 0,
\end{equation}
using the coordinates ${\tilde X}_i$. Here the Skyrmion gives rise 
to 

\begin{equation}
U^{\gamma_5} = \exp \left( i f(\mu(R)) \gamma_5 \tau_i \frac{X_i}{R} 
\right),
\end{equation}
and the same equation holds in the coordinate system ${\tilde X}_i$.

\section{Solving the Dirac Equation for $G = 0$}
\label{DiracSolution}

In order to solve the Dirac equation we introduce spherical polar
coordinates for equation (\ref{Dirac1}) and (\ref{Dirac2}). 
In terms of the coordinates $x = (x_1,x_2,x_3,w)$  in ${\mathbb R}^4$, the polar
coordinates for the $3$-sphere of radius $1$ can be written as
\begin{equation}
x = \left( \sin \mu \sin \theta \cos \phi,
\sin \mu \sin \theta \sin \phi,
\sin \mu \cos \theta,
\cos \mu
\right).
\end{equation}
The angles $(\mu,\theta,\phi)$ can now be expressed in terms of $X_i$ as
\begin{eqnarray}
\nonumber
\mu &=& \left\{
\begin{array}{ll}
\arctan \frac{2R}{R^2-1} +\pi & {\rm for}~~~ R^2<1, \\
\frac{\pi}{2} & {\rm for}~~~ R^2 = 1,\\
\arctan \frac{2R}{R^2-1}      & {\rm for}~~~ R^2>1,
\end{array} 
\right. \\
\nonumber
\theta &=& \arctan \frac{\sqrt{X_1^2 + X_2^2}}{X_3}, \\
\phi &=& \arctan \frac{X_2}{X_1}.
\end{eqnarray}
In terms of the  ${\tilde X}_i $, the relations for
$\theta$ and $\phi$ are identical to the ones above, 
and for the $\mu$ coordinate we obtain
\begin{eqnarray}
\mu &=& \left\{
\begin{array}{ll}
\arctan \frac{2 {\tilde R}}{1-{\tilde R}^2}  & {\rm for}~~~ 
{\tilde R}^2<1, \\ 
\frac{\pi}{2} & {\rm for}~~~ R^2 = 1,\\
\arctan \frac{2 {\tilde R}}{1-{\tilde R}^2} + \pi     & {\rm for}~~~
{\tilde R}^2>1.
\end{array} 
\right. 
\end{eqnarray}
Equation (\ref{Dirac1}) gives rise to the following Dirac equation in polar
coordinates: 
\begin{equation}
\label{Dpolar1}
\left( i \gamma^0 \partial_t + 
i \gamma^i \left(
e_{\mu i} \partial_\mu -
\tfrac{
e_{\theta i} \partial_\theta
}{\sin \mu} 
- 
\tfrac{
e_{\phi i} \partial_\phi
}{\sin \mu \sin \theta} 
+
\tfrac{ 
e_{\mu i} 
\sin \mu}{1 - \cos \mu} 
\right)
- g~U^{\gamma_5} 
\right)
\psi = 0,
\end{equation}
where $\psi = \psi (\mu,\theta, \phi,t)$ and 
the unit vectors $(e_\mu,e_\theta,e_\phi)$ are given by
\begin{eqnarray}
\label{unitvectors}
e_\mu = \left(
\begin{array}{c}
\sin \theta \cos \phi \\
\sin \theta \sin \phi \\
\cos \theta
\end{array}
\right),\
e_\theta = \left(
\begin{array}{c}
\cos \theta \cos \phi \\
\cos \theta \sin \phi \\
- \sin \theta
\end{array}
\right)\
{\rm and}\
e_\phi = \left(
\begin{array}{c}
- \sin \phi \\
\cos \phi \\
0
\end{array}
\right).
\end{eqnarray}
$U^{\gamma_5}$ can be written as 
$U^{\gamma_5} = \cos f(\mu) + i \gamma_5 {\bf e_\mu} \cdot \vectau
\sin f(\mu)$. Similarly,
equation (\ref{Dirac2}) give rise to 
\begin{equation}
\label{Dpolar2}
\left( i \gamma^0 \partial_t + 
i \gamma^i \left(
e_{\mu i} \partial_\mu +
\tfrac{e_{\theta i} \partial_\theta}{\sin \mu} + 
\tfrac{e_{\phi i} \partial_\phi}{\sin \mu \sin \theta} -
\tfrac{e_{\mu i} \sin \mu}{1 + \cos \mu} 
\right)
- g~U^{\gamma_5}
\right)
{\tilde \psi}  = 0.
\end{equation}
The spin connection term could be removed by setting $\psi =
\frac{1}{1 - \cos \mu} \Phi$ and 
${\tilde \psi} = \frac{1}{1+\cos \mu} {\tilde \Phi}$, respectively. 
However, it is
easier to discuss the regularity of the solution in terms of the
original fields.

We adopt the standard representation of the
$\gamma$-matrices where $\gamma^0$ is diagonal. This is convenient
because the Dirac equation is invariant under parity. 
\begin{eqnarray}
\gamma^0 =
\left(
\begin{array}{cc}
I_2 & 0 \\
0 & -I_2
\end{array}
\right),\
\gamma^i =
\left(
\begin{array}{cc}
0 & \sigma_i \\
-\sigma_i & 0
\end{array}
\right),\
{\rm and}\
\gamma_5 =
\left(
\begin{array}{cc}
0 & I_2 \\
I_2 & 0
\end{array}
\right).
\end{eqnarray}
Now, the spinor transformation $\rho$ in (\ref{rhospin}) can be
evaluated explicitly: 
\begin{equation}
\label{trafo}
{\tilde \psi} \left({\tilde X_i}\right) =
\left(
  \begin{array}{cc}
     -i \sigma_j {\hat X}_j & 0 \\
     0 & -i \sigma_j {\hat X}_j 
  \end{array}
\right) 
\psi \left(X_i \right).
\end{equation}
With this choice of $\gamma$-matrices, 
we can derive the differential equation for the stationary solutions
of the Dirac equation by setting $\psi(X_i,t) = {\rm e}^{i E t}
\psi(X_i)$: 
\begin{eqnarray}
\label{eq1}
E \psi =
\left(
\begin{array}{cc}
g \cos f(\mu) 
& \vecs \cdot {\bf p} + ig {\bf e_\mu} \cdot \vectau \sin f(\mu) \\ 
  \vecs \cdot {\bf p} - ig {\bf e_\mu} \cdot \vectau \sin f(\mu) 
& - g \cos f(\mu)
\end{array}
\right) \psi,
\end{eqnarray}
where
\begin{equation}
\vecs \cdot {\bf p} = -i \left( {\bf e_\mu} \cdot \vecs \left(\partial_\mu
+ \tfrac{\sin \mu}{1-\cos \mu}\right) - 
\tfrac{1}{\sin \mu} {\bf e_\theta} \cdot \vecs \partial_\theta 
- \tfrac{1}{\sin \mu \sin \theta} 
{\bf e_\phi} \cdot \vecs \partial_\phi
\right).
\end{equation}
The components of the matrix in equation (\ref{eq1}) commute with the
total angular momentum  operator 
${\bf G =} {\bf L} + {\bf S} + {\bf I}$ where ${\bf L}$ is the orbital
angular momentum operator, the spin operator is ${\bf S} =
\frac{1}{2} \vecs$ and the isospin operator is ${\bf I} = \frac{1}{2}
\vectau$. In polar coordinates, the orbital angular momentum ${\bf L}$
is given by  
\begin{eqnarray}
\nonumber
L_x &=& i\left(\sin \phi \frac{\partial}{\partial \theta} 
      + \cot \theta \cos \phi \frac{\partial}{\partial \phi}\right), \\
\nonumber
L_y &=& i\left(- \cos \phi \frac{\partial}{\partial \theta} 
      + \cot \theta \sin \phi \frac{\partial}{\partial \phi}\right), \\
L_z &=& -i \frac{\partial}{\partial \phi}.  
\end{eqnarray}
Equation (\ref{eq1}) is also invariant under
parity ${\hat P}$ with
\begin{equation}
{\hat P}~ \psi\big(X_i \big) = \gamma_0~ \psi \big( - X_i \big)\
{\rm and}\ {\hat P}~ X_i~ {\hat P}^{-1} = - X_i. 
\end{equation}
It is worth discussing the parity operator ${\hat P}$ in more
detail. ${\hat P}$ acts on the $\gamma$-matrices and the partial
derivatives as follows:
\begin{eqnarray} 
\begin{array}{ccc}
{\hat P}~ \gamma_0~ {\hat P}^{-1} = \gamma_0 & {\rm and} & 
{\hat P}~ \gamma_i~ {\hat P}^{-1} = - \gamma_i, \\
{\hat P}~ \partial_0~ {\hat P}^{-1} = \partial_0 & {\rm and} &
{\hat P}~ \partial_i~ {\hat P}^{-1} = - \partial_i. 
\end{array}
\end{eqnarray}
Using the spinor transformation $\rho$ in (\ref{rhospin}) we can show that 
\begin{equation}
{\hat P}~ {\tilde \psi}\big({\tilde X}_i\big) = 
 - \gamma_0~ {\tilde \psi}\big({\tilde X}_i\big).
\end{equation}
Therefore, the parity operator ${\hat {\tilde P}}$ which is defined
for the coordinates ${\tilde X}_i$ differs from 
${\hat P}$ by a minus sign.\footnote{The origin of this sign difference is that 
the charts do not form an orientation atlas.}

The $G^P = 0^+$ positive parity spin isospin spinors are given by  
\begin{eqnarray}
\label{0+spinor}
\psi \left(\mu,\theta,\phi \right) = 
\left(
\begin{array}{c}
\left(1 - \cos \mu\right)^{\frac{1}{2}} G(\cos \mu) \sigma_2 \\
i \left(1 - \cos \mu\right)^{\frac{1}{2}} F(\cos \mu) 
\left({\bf e}_\mu \cdot \vecs \right) \sigma_2
\end{array}
\right).
\end{eqnarray}
This gives rise to the following system of first order differential
equations:
\begin{eqnarray}
\label{general1}
\frac{{\rm d}}{{\rm {d}} \mu}
\left(
\begin{array}{c}
F \\
G
\end{array}
\right) = \left(
\begin{array}{cc}
\frac{1 - 3 \cos \mu}{2\sin \mu} - g \sin f & E - g \cos f \\
- \left(E + g \cos f \right) & 
        g \sin f
        -\frac{3 \left(1 + \cos \mu \right)}{2\sin \mu} 
\end{array}
\right) \left(
\begin{array}{c}
F \\
G
\end{array}
\right),
\end{eqnarray}
where $F$, $G$ and $f$ are functions of $\mu$.
Note that the ${G}^P = {0}^-$ spinor
\begin{eqnarray}
\psi \left(\mu,\theta,\phi \right) = 
\left(
\begin{array}{c}
i \left(1 - \cos \mu\right)^{\frac{1}{2}} F(\cos \mu) 
\left({\bf e}_\mu \cdot \vecs \right) \sigma_2 \\
\left(1 - \cos \mu\right)^{\frac{1}{2}} G(\cos \mu) \sigma_2 
\end{array}
\right)
\end{eqnarray}
gives rise to the same equation with $g$ replaced by $-g$.

In ${\tilde X}_i$-coordinates positive and negative parity are
interchanged such that the ${0}^+$ is given by
\begin{eqnarray}
{\tilde \psi}\left(\mu,\theta,\phi \right) = 
\left(
\begin{array}{c}
i \left(1 + \cos \mu \right)^{\frac{1}{2}} {\tilde F}(\cos \mu) 
\left({\bf e}_\mu \cdot \vecs \right) \sigma_2 \\
\left(1 + \cos \mu\right)^{\frac{1}{2}} {\tilde G}(\cos \mu) \sigma_2 
\end{array}
\right).
\end{eqnarray}
This gives rise to 
\begin{eqnarray}
\label{general2}
\frac{{\rm d}}{{\rm {d}} \mu}
\left(
\begin{array}{c}
{\tilde F} \\
{\tilde G}
\end{array}
\right) = \left(
\begin{array}{cc}
- \frac{1 + 3 \cos \mu}{2 \sin \mu} + g \sin f & E + g \cos f \\
- \left(E - g \cos f \right) & 
      \frac{3 \left(1 - \cos \mu \right)}{2 \sin \mu} - g \sin f
\end{array}
\right) \left(
\begin{array}{c}
{\tilde F} \\
{\tilde G}
\end{array}
\right).
\end{eqnarray}
Again the ${0}^-$ spinor
\begin{eqnarray}
{\tilde \psi} \left(\mu,\theta,\phi \right) = 
\left(
\begin{array}{c}
\left(1 + \cos \mu\right)^{\frac{1}{2}} {\tilde G}( \cos \mu) \sigma_2 \\
i \left(1 + \cos \mu \right)^{\frac{1}{2}} {\tilde F}(\cos \mu) 
\left({\bf e}_\mu \cdot \vecs \right) \sigma_2 
\end{array}
\right)
\end{eqnarray}
gives rise to the same equation with $g$ replaced by $-g$.

Equation 
(\ref{general1}) can be transformed into a second order
differential equation for $G(u)$ where $u= \cos \mu$. Note that
since $\mu \in [ 0, \pi ]$ it is consistent to set $\sin \mu = \sqrt{1
- \cos^2 \mu}$. 
\begin{equation}
\nonumber
\left( \left(1-u^2 \right) \frac{{\rm d}^2}{{\rm d}u^2} 
- \left(4 u +1+ \frac{g~f^\prime \sin f \sqrt{1-u^2}}{E + g \cos f}
\right) \frac{{\rm d}}{{\rm d}u} +
E^2 -g^2   - \tfrac{9}{4}
+ \frac{2g \sin f}{\sqrt{1-u^2}} 
\right.  
\end{equation}
\begin{equation}
\label{OD1}
\left.
+ \frac{3 g f^{\prime} \sin f}{2 \left(E + g \cos f \right)} 
\frac{(1+u)}{\sqrt{1-u^2}} 
- \frac{g^2 f^\prime \sin^2 f}{E + g \cos f} - g f^\prime \cos f
\right) G(u) = 0, 
\end{equation}
and similarly equation 
(\ref{general2}) can be transformed into
\begin{equation}
\nonumber
\left( \left(1-u^2 \right) \frac{{\rm d}^2}{{\rm d}u^2} 
- \left(4 u -1 - \frac{g~f^\prime \sin f \sqrt{1-u^2}}{E - g \cos f}
\right)  \frac{{\rm d}}{{\rm d}u} 
+E^2 -g^2 - \tfrac{9}{4}
+\frac{2g \sin f}{\sqrt{1-u^2}} 
\right.  
\end{equation}
\begin{equation}
\label{OD2}
\left.
+ \frac{3 g f^\prime \sin f}{2 \left(E - g \cos f \right)} 
\frac{(1-u)}{\sqrt{1-u^2}} 
- \frac{g^2 f^\prime \sin^2 f}{E - g \cos f} + g f^\prime \cos f
\right) {\tilde G}(u) = 0. 
\end{equation} 
It is not easy to transform equation (\ref{OD1}) and (\ref{OD2}) into
each other. In fact, 
due to the spinor transformation (\ref{trafo}) 
${\tilde G}(u)$ and $F(u)$ are related via
\begin{equation}
\label{FGtilde} 
F(u) = \sqrt{\frac{1+u}{1-u}}{\tilde G}(u).
\end{equation}
Also note that $F(u)$ is given by
\begin{equation}
\label{F}
F(u) = \frac{1}{E + g \cos f}\left(
\left(
-\frac{3 (1 + u)}{2 \sqrt{1-u^2}} + g \sin f
\right) G(u) 
- \sqrt{1 - u^2} \frac{{\rm d}}{{\rm d}u} G(u)
\right),
\end{equation} 
and similarly ${\tilde F}(u)$ can be written as
\begin{equation}
\label{tildeF}
{\tilde F}(u) = \frac{1}{E - g \cos f}\left(
\left(
\frac{3(1 - u)}{2 \sqrt{1-u^2}} - g \sin f
\right) {\tilde G}(u) 
- \sqrt{1 - u^2} \frac{{\rm d}}{{\rm d}u} {\tilde G}(u)
\right).
\end{equation}

\section{Explicit Solutions of the Dirac equation}

In this section we solve the Dirac equation which has been derived in the 
previous section for various shape functions $f(\mu)$. However, first we 
review some basic facts about ordinary differential equations 
and regular singular points, following \cite{Whittaker:1927}.

\subsection{Differential Equations and their Singular Points}
\label{DifferentialEquations}

Any linear second order differential equation can be brought into the
standard form 
\begin{equation}
\label{diffeq}
y^{\prime \prime}(u) + P(u) y^\prime(u) + Q(u) y(u) = 0.
\end{equation}
If $P(u)$ and $Q(u)$ are analytic functions in the neighbourhood of
$u=u_0$ then $u = u_0$ is called a regular point. 
In this case there are locally two linearly independent solutions
$y_1(u)$ and $y_2(u)$ such that any solution of (\ref{diffeq}) is
a linear combination of $y_1(u)$ and $y_2(u)$. Moreover,
$y(u)$ is an analytic function, {\it i.e.} it can be written as
\begin{equation}
y(u) = \sum\limits_{n=0}^\infty a_n (u-u_0)^n.
\end{equation}
If either $P(u_0)$ or $Q(u_0)$ is singular then $u=u_0$ is called a
singular point. In the special case that 
\begin{eqnarray}
P(u) &=& \frac{p(u)}{u-u_0}, \\
Q(u) &=& \frac{q(u)}{(u-u_0)^2},
\end{eqnarray}
where $p(u)$ and $q(u)$ are analytic functions, {\it i.e.} 
\begin{equation}
p(u) = p_0 + p_1 (u-u_0) + \dots\ {\rm and}~~~
q(u) = q_0 + q_1 (u-u_0) + \dots,
\end{equation}
the point $u=u_0$ is
called a {\it regular singular point}. 
Equations which only have regular or regular singular points are known
as Fuchsian differential equations.
Equation (\ref{diffeq}) can then be solved by the following
ansatz:  
\begin{equation}
\label{eqansatz}
y(u) = (u-u_0)^\rho \sum\limits_{n=0}^\infty a_n (u-u_0)^n.
\end{equation}
The exponent $\rho$ can be calculated by solving the so-called
{\it indicial equation} 
\begin{equation}
\label{indicial}
\rho^2 + (p_0 - 1) \rho + q_0 = 0,
\end{equation}
which is derived by inserting the
ansatz (\ref{eqansatz}) into (\ref{diffeq}).
The two solutions $(\rho_1,\rho_2)$ of the indicial equation
(\ref{indicial}) are called the {\it exponents} of the regular singular
point $u_0$. Our convention will be that $\rho_1 \ge \rho_2$.
If the difference of the exponents $s = \rho_1 - \rho_2$ is
not a positive integer or zero, then the ansatz (\ref{eqansatz}) 
gives rise to two linearly independent 
solutions of (\ref{diffeq}) which are uniformly
convergent in some neighbourhood of $u_0$. Note, however, that the
solution diverges at $u_0$ if its exponent $\rho$ is negative.

Let $s = \rho_1 - \rho_2$ be a positive integer or zero. 
This situation is slightly more subtle because the two solutions can
become linearly dependent, and this could lead to further logarithmic
terms.  
The solution $y_1(u)$ with exponent $\rho_1$ is always uniformly 
convergent in some neighbourhood of $u_0$. 
Therefore, we reduce equation (\ref{diffeq}) to a
differential equation of first order by setting $y(u) = y_1(u)
z(u)$. This gives rise to the following equation for $z(u)$:
\begin{equation}
\label{difeqz}
(u-u_0)^2 z^{\prime \prime}(u) + \left(
2 (u-u_0)^2 \frac{y_1^\prime(u)}{y_1(u)} + (u-u_0) p(u) 
\right) z^\prime(u) = 0.
\end{equation}
This equation can be solved by integration and separation of variables,
and we obtain
\begin{eqnarray}
\nonumber
z(u) &=& A + B_1 \int\limits_{}^{u} 
\frac{1}{\left( y_1(v)\right)^2} 
\exp
\left( -\int\limits_{}^{v} \frac{p(w)}{w-u_0} {\rm d} w
\right) {\rm d}v, \\
\nonumber
&=& A + B_2 \int\limits_{}^u \frac{v-u_0}{\left( y_1(v)\right)^2}
\exp \left(
- p_1 (v-u_0) - \tfrac{1}{2} p_2 (v-u_0)^2 -\dots
\right) {\rm d}v, \\
\label{ggss}
&=& A + B_3 \int\limits_{}^{u} (v - u_0)^{-p_0 -2\rho_1} g(v) {\rm d}v, 
\end{eqnarray}
where $g(u)$ is an analytic function in a suitable domain, and $A$ and 
$B_i$ are constants. A short calculation shows that 
\begin{equation}
-p_0 -2\rho_1 = -s-1.
\end{equation}
Let
\begin{equation}
\label{gs}
g(u) = 1 + \sum\limits_{n=1}^{\infty} g_n (u-u_0)^n.
\end{equation}
Then for $s \neq 0$ the most general solution is 
\begin{equation}
y(u) = A y_1(u) + B \left(g_s y_1(u) \log(u - u_0) + {\tilde y}_2(u)
\right),
\end{equation}
where $A$ and $B$ are arbitrary constants, and $y_1(u)$ is the
solution of the form (\ref{eqansatz}).
The constant $g_s$ and the
function ${\tilde y_2}(u)$ are completely determined by the
differential equation (\ref{difeqz}). 
The function ${\tilde y}_2(u)$ is given by
\begin{equation}
{\tilde y}_2(u) = (u-u_0)^{\rho_2} \left( -\frac{1}{s} + 
\sum\limits_{n=1}^\infty h_n(u-u_0)^n \right),
\end{equation}
the coefficients $h_n$ being constants. In the special case that $g_s
= 0$ no logarithms occur, {\it i.e.} the two solutions $y_1(u)$ and
$y_2(u)$ are linearly independent.

If $s=0$ the corresponding form of the solution is 
\begin{equation}
\label{log}
y(u) = A y_1(u) + B \left(y_1(u) \log(u - u_0) 
+ (u-u_0)^{\rho_2} \sum\limits_{n=1}^\infty h_n(u-u_0)^n
\right).
\end{equation}

It is also possible to discuss the singularity at infinity. The
transformation $v = \frac{1}{u}$ maps the point at infinity to
the origin. If we apply this transformation to equation
(\ref{diffeq}) we obtain a new differential equation in terms of $v$. 
If the new equation has a regular or a regular singular point at the
origin $v=0$ then (\ref{diffeq}) has a regular or regular singular
point at infinity. 

If a Fuchsian differential equation has $n$ finite regular singular
points $u=u_i$ with exponents $\rho_{1,2}^{(i)}$ and a regular
singular point at infinity with exponents $\rho_{1,2}^{(\infty)}$ 
then the following equation holds:
\begin{equation}
\label{erelation}
\rho_1^{(\infty)} + \rho_2^{(\infty)} + \sum\limits_{i=1}^{n} \left(
\rho_1^{(i)} + \rho_2^{(i)} \right) = n-1.
\end{equation}
This is a useful consistency check for the exponents.

\subsection{Fermions coupled to a Constant Skyrme Field}
\label{ConstantSkyrmion}

We are interested in the behaviour of fermions in the presence of a
background Skyrmion. 
The Skyrmion introduces a space-dependent
complex mass for the fermions. Therefore, it is instructive to
consider first a massive fermion on $S^3$. 
This corresponds to $f(\mu) \equiv 0$ so that the fermion mass is
$g$. This example can be solved explicitly in terms of hypergeometric
functions. 
In the following sections we proceed to more complicated
configurations. In Sect. \ref{HedgehogSkyrmion}, we consider the
spherically symmetric shape function $f(\mu) = \mu$ which can also be
solved explicitly, but in a less standard way. In Sect.
\ref{Generalization}, the conformal shape function (\ref{conformal}) 
is discussed. In
this case, the differential equation can no longer be solved
explicitly. 
We present numerical algorithms in Sect. \ref{NumericalAlgorithm}.

Setting $f(\mu)=0$, equation (\ref{OD1}) can be written as
\begin{equation}
\label{OD1f0}
\left(\left(1-u^2\right)\frac{{\rm d}^2}{{\rm d} u^2} - 
\left(4 u+1 \right) \frac{{\rm
d}}{{\rm d}u} + e^2 - \tfrac{9}{4}
\right) G(u) = 0,
\end{equation}
where $e = \sqrt{E^2 -g^2}$.

Equation (\ref{OD1f0}) has three regular singular points, one at
$u=1$, one at $u= -1$ and one at 
$u= \infty $ with exponents $\rho^{(1)}_{1,2}=(0,-\frac{3}{2})$, 
$\rho^{(-1)}_{1,2}=(0,-\frac{1}{2})$ and 
$\rho^{(\infty)}_{1,2}=(\frac{3}{2}+e,\frac{3}{2}-e)$ 
respectively. Note that relation (\ref{erelation}) between the exponents
is satisfied, {\it i.e.} the sum of the exponents is
equal to $1$.
In this example we will show that the solution is non-singular on
$S^3$ if its exponents are $\rho^{(-1)}=0$ and $\rho^{(1)}=0$.

Equation (\ref{OD1f0}) can be solved explicitly:
\begin{eqnarray}
\nonumber
G(u) &=& 
c_1 {\cal F} \left(
\tfrac{3}{2} + e,\tfrac{3}{2}-e;\tfrac{3}{2};\tfrac{1}{2} \left(1+u \right)
\right) \\
&+& c_2 \left(1+u\right)^{-\frac{1}{2}} 
{\cal F} \left(
1+e,1-e;\tfrac{1}{2};\tfrac{1}{2} \left(1+u \right)
\right),
\end{eqnarray}
where ${\cal F}(a,b;c;z)$ is a hypergeometric function, {\it e.g.} see
\cite[p. 556]{Abramowitz:1972}.
The hypergeometric function is non-singular for $0 \le z<1$ and is
normalized at $z=0$, {\it i.e.} ${\cal F}(a,b;c;0)=1$. It may be
singular at $z=1$, corresponding to $u= 1$, which is the north pole. 
As $G(u)$ has to be non-singular at the south pole, $u = -1$, we
have to set $c_2=0$. 

Since $G(u)$ is not well defined at $u=1$ we have to express $G(u)$ in
terms of ${\tilde G}(u)$ and calculate ${\tilde G}(1)$.
Using the transformation (\ref{trafo}) from ${\tilde \psi}$ to $\psi$
we obtain 
\begin{eqnarray}
\nonumber
G(u) &=& 
\left( \frac{1+u}{1-u} \right)^{\frac{1}{2}}
\left( -i \sigma \cdot e_\mu~ i \sigma \cdot e_\mu 
{\tilde F}(u) \right), \\
&=& \left( \frac{1+u}{1-u} \right)^{\frac{1}{2}}{\tilde F}(u),
\end{eqnarray}
which is the analogue of equation (\ref{FGtilde}).
Setting $f(\mu) = 0$, in equation (\ref{OD2}) we can
derive a similar equation for ${\tilde G(u)}$ which has exponents
$\rho^{(-1)}_{1,2}=(0,-\tfrac{3}{2})$,
$\rho^{(1)}_{1,2}=(0,-\tfrac{1}{2})$ and
$\rho^{(\infty)}_{1,2}=(\tfrac{3}{2}+e,\tfrac{3}{2}-e)$. The solution
of this equation can again be expressed in terms of hypergeometric
functions: 
\begin{eqnarray}
\nonumber
{\tilde G}(u) &=& 
{\tilde c}_1 {\cal F}\left( 
\tfrac{3}{2} + e,\tfrac{3}{2}-e;\tfrac{3}{2};\tfrac{1}{2} \left(1-u \right)
\right) \\
&+& {\tilde c}_2 \left(1-u\right)^{-\frac{1}{2}} {\cal F} \left(
1+e,1-e; \tfrac{1}{2};\tfrac{1}{2} \left(1-u \right)
\right).
\end{eqnarray}
This is only non-singular at $u=1$ if we set ${\tilde c}_2 =0$. 
Assuming that $E-g \neq 0$, ${\tilde F}(u)$ can be calculated from 
equation (\ref{tildeF}):
\begin{equation}
{\tilde F}(u) = \frac{1}{E-g} \left(
- \frac{3(1-u)}{2 \sqrt{1-u^2}}~ {\tilde G}(u) 
+ \sqrt{1-u^2} \frac{{\rm d}}{{\rm d} u} {\tilde G}(u) 
\right).
\end{equation}
Therefore, $G(u)$ can be written as
\begin{equation}
\label{G(1)nonsing}
G(u) = \frac{1}{E-g} \left(
(1+u) \frac{{\rm d}}{{\rm d} u} {\tilde G}(u) 
-\tfrac{3}{2} {\tilde G}(u)
\right).
\end{equation}
${\tilde G}(u)$ has to be non-singular at $u=1$.
Equation (\ref{G(1)nonsing} then implies that $G(u)$ is also non-singular 
at $u=1$. Therefore, the regular solution $G(u)$ has the exponents
$\rho^{(-1)}=0$ and $\rho^{(1)}=0$. Note that this argument only relies on 
the local behaviour and explicit solutions are not needed.
As we shall see, in the generic case, the exponents at the singular
points do not depend on the chosen shape function, so that our
analysis is also valid for general shape functions. 
Problems can occur if additional singular
points coalesce with the singular points at $u=\pm 1$ because this changes 
the exponents.\footnote{In the following section we will also discuss the 
case that further singular points are inside the interval $(-1,1)$.}
In the generic case, the regular solution can be calculated by
imposing that the solution can be expanded as a power series at the
north and at the south pole. Later, this will be the basis for our
numerical algorithm. Now, we impose this condition in order to calculate 
the spectrum of equation (\ref{OD1f0}).

Using formula (15.3.6) in \cite{Abramowitz:1972} we can express  $G(u)$ in
terms of hypergeometric functions at the second singular
point.\footnote{
Equation (\ref{hypid}) only holds if $|{\rm arg}(1-z)| < \pi$. 
However, this inequality is satisfied because  
$(1-z) = \frac{1}{2}(1+\cos \mu) \geq 0$.}
\begin{eqnarray}
\nonumber
G(u) &=& c_1 {\cal F} \left(\tfrac{3}{2}+ e,\tfrac{3}{2}-e;\tfrac{3}{2};
                 \tfrac{1}{2} \left(1+u \right) \right),
\\ 
\label{hypid}
     &=& c_1 \left( 
    \frac{\Gamma{(\tfrac{3}{2})} \Gamma{\left(-\tfrac{3}{2}\right)}}
    {\Gamma{(e)}\Gamma{(-e)}}
    {\cal F} \left(\tfrac{3}{2}+e,\tfrac{3}{2}-e;\tfrac{5}{2};
           \tfrac{1}{2}\left(1-u \right) \right) 
\right.   \\
\nonumber
&+&
\left. 
   \left(\frac{1-u}{2}\right)^{-\frac{3}{2}}
     \frac{\Gamma{\left(\tfrac{3}{2}\right)}\Gamma{\left(\tfrac{3}{2}\right)}}
     {\Gamma{\left(\tfrac{3}{2}-e\right)}\Gamma{\left(\tfrac{3}{2}+e\right)}}
     {\cal F} \left(-e,e;-\tfrac{1}{2};\tfrac{1}{2}\left(1-u \right) \right)
\right).
\end{eqnarray}
The last term has to vanish, otherwise ${\tilde F}(u)$ would be singular
at the north pole. Therefore, either 
$\Gamma(\frac{3}{2} + e)$ or $\Gamma(\frac{3}{2}-e)$ has to have a
pole. Since $e \geq 0$, we obtain $e = N + \frac{3}{2}$, and
the spectrum is therefore
\begin{equation}
\label{specf=0}
E = \pm \sqrt{g^2 + \left(N+\tfrac{3}{2}\right)^2},~~~ {\rm
where}~~~ N = 0, 1, 2, \dots  
\end{equation}
With this condition the eigenfunctions $G_N(u)$ reduce to
Jacobi Polynomials (\cite{Abramowitz:1972}, formula 15.4.6).

\begin{figure}[!htb]
\begin{center}
\includegraphics[height=125mm,angle=270]{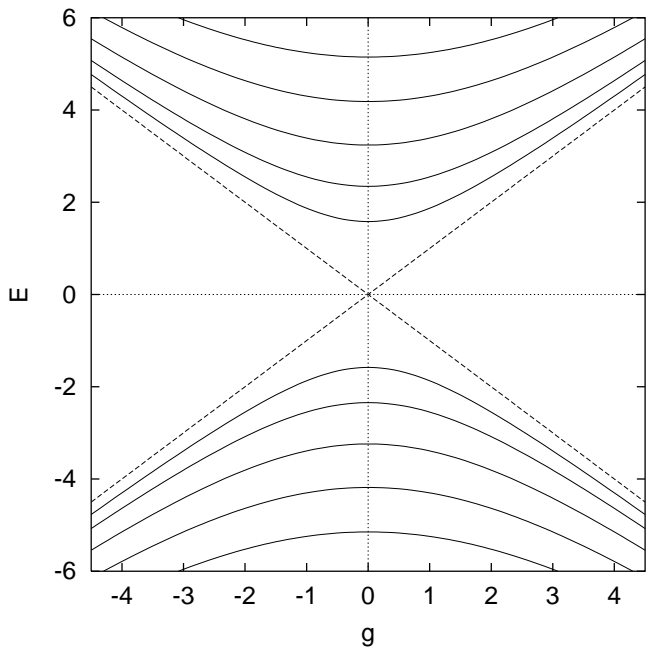}
\caption{The energy levels $E_n$ as a function of the coupling
constant $g$ for $f(\mu) = 0$. $E = \pm g$ is shown as dashed lines. 
\label{fermion1}}
\end{center}
\end{figure}

The spectrum is displayed in figure \ref{fermion1}.
It is symmetric with respect to $E=0$. The importance
of this result lies in the fact that for $g=0$ the energy $E_N$ and the
corresponding solutions $G_N(u)$ are the same for all possible shape
functions $f(\mu)$. 
Therefore, these solutions can be used as starting values
for the relaxation method, which will be described in Sect.
\ref{Numerics}. As the radius of the $3$-sphere tends to infinity the spectrum 
tends to the usual spectrum of the Dirac equation, which is continuous for 
$E^2 \ge g^2$. 
Also note that if we assume that $G(u)$ and its first and second 
derivative remain finite as $g$ tends to infinity then the energy $E$ has 
to be of order $g$ in this limit, so $\pm g$ is the leading order 
asymptotic for $E$. Therefore, we displayed the lines $E=\pm g$ in figure 
\ref{fermion1}. 
As we shall see in Sect. \ref{Numerics} the region $E^2 < g^2$ also plays 
a special role for our numerical algorithm.

Another way of deriving the spectrum (\ref{specf=0}) is to note
that if the exponents at $u= \pm 1$ both vanish then $G(u)$ is an 
integral function, that is it has only one singular point at infinity. 
The only integral functions which do not possess an essential singular 
point at infinity are the polynomials, {\it e.g.} 
\cite[p. 106]{Whittaker:1927}.
Therefore, one of the exponents at infinity is equal to $-N$ for $N = 
0,1,\dots$. Since the exponents at
infinity are $\rho^{(\infty)}_{1,2}=(e+\frac{3}{2},-e+\frac{3}{2})$ 
we obtain the same spectrum as before. Note however, that since the 
exponents $\rho^{(\infty)}_{1,2}$ differ by an integer, we have to 
check in this approach that the solution is indeed a polynomial, {\it 
i.e.} $g_s$ in equation (\ref{gs}) vanishes.
This spectrum has also been calculated by other authors, {\it 
e.g.} \cite{Sen:1986dc, Carmeli:1985kd}. 

\subsection{Fermions coupled to the Skyrme Field $f(\mu) = \mu$}
\label{HedgehogSkyrmion}

In this section we construct the solutions of equation (\ref{OD1})
for $f(\mu) = \mu$. Configurations with $f(\mu) \neq 0$ are called 
``hedgehog configurations'' because at each point the vector in isospin
space is proportional to the coordinate vector. For $f(\mu) =\mu$ the
map from the base $S^3$ to the target $S^3$ is even an isometry and 
possesses $O(4)$-symmetry. Setting $f(\mu) = \mu$ in
equation (\ref{OD1}) we obtain 
\begin{equation}
\label{G(u)k=1}
\left(1-u^2\right)\left(E+gu\right) \frac{{\rm d}^2 G}{{\rm d} u^2} 
- \left(\left(4u+1\right)\left(E+gu\right) + g\left(1-u^2\right)\right) 
\frac{{\rm d} G}{{\rm d} u} + 
\end{equation}
\begin{equation}
\nonumber
\left( \left(E+gu \right) \left(E^2 - g^2 + 2g -\tfrac{9}{4}-gu\right) 
+\tfrac{3}{2}\, g \left(1+u \right) - g^2 \left(1-u^2 \right)\right) G = 0.
\end{equation}
This is a homogeneous differential equation of Fuchsian type with four
singular points. They are located at $u= \pm 1$, $u= -\frac{E}{g}$ and
$u= \infty$. Assuming $E \neq \pm g$ the exponents are
\begin{eqnarray}
\nonumber
\rho_{1,2}^{(1)}  &=& \left(0,-\tfrac{1}{2} \right),  \\
\nonumber
\rho_{1,2}^{(-1)} &=& \left(0,-\tfrac{3}{2} \right),   \\
\nonumber
\rho_{1,2}^{(-\frac{E}{g})} &=& \left(2,0 \right),    \\
\label{exk=1}
\rho_{1,2}^{(\infty)} &=& \left(
1 \pm \tfrac{1}{2} \sqrt{1-4E+4E^2+ 8g - 4g^2}
\right).
\end{eqnarray}
For $E=g$ the singular points at $u=-1$ and $u = - \frac{E}{g}$
coalesce, and therefore, the exponents at  $u=-1$ become $\rho_{1,2}^{(-1)} =
(\frac{1}{2},0)$ and the exponent at infinity is $\rho_{1,2}^{(\infty)} =
(1 \pm \frac{1}{2} \sqrt{1+4g})$. 
The relation for the exponents (\ref{erelation}) is satisfied.
For $E=-g$ the singular points at $u=1$ and $u = -\frac{E}{g}$
coalesce. This changes the exponents at $u=1$ to $\rho_{1,2}^{(1)} =
(-1,-\frac{3}{2})$. In this case, there is an essential singular point
at infinity and the equation is no longer Fuchsian.

Since we are interested in solutions that can be expanded as a power series
around $u= \pm 1$, it is natural to look for polynomial solutions.
The singular point at  $u=-\frac{E}{g}$ might impose a
constraint. In order to verify whether there is a logarithmic term we
have to calculate the  coefficient $g_s$ in equation (\ref{gs}). Here
$s=2$. After a standard calculation we obtain $g_2 =
0$. Therefore, there are two independent power series at the point $u
= -\frac{E}{g}$. This imposes no further restriction, since we do not
have to exclude any of these two solutions. This implies that $G(u)$ is 
again an integral function and we can calculate the energy spectrum using 
the exponents at infinity. Again, the exponents $\rho^{(\infty)}_{1,2}$ 
differ by an integer, and we have to verify that the solutions are 
indeed polynomials. It turns out that the only problem occurs for 
$\rho^{(\infty)}_{2} = 0$.

In the following, we derive the recurrence relation for these
polynomials. Yet, it is instructive to consider first the solution where
$G(u)$ is a constant $a_0$, {\it i.e.} $\rho^{(\infty)}_{2} = 0$. Then we 
can solve equation (\ref{exk=1}) for $E$ and obtain $E=g-\frac{1}{2}$ and 
$E=\frac{3}{2} - g$. Setting $G(u) = a_0$ in equation (\ref{G(u)k=1}), we 
obtain 
\begin{equation}
\left(E+g-\tfrac{3}{2} \right)
\Big(
g \left( E-g+ \tfrac{1}{2} \right) \left( u+1 \right) +
\left( E-g+\tfrac{3}{2} \right) \left( E-g \right) 
\Big) a_0 = 0.
\end{equation}
It is easy to see that this is not a solution for $E = g - \frac{1}{2}$.
$G(u) = a_0$ is only a solution  for arbitrary $g$ if $E=\frac{3}{2}-g$. 
This solution is very interesting because its spectrum crosses $E=0$ for 
$g = \frac{3}{2}$.

Now we derive the general solution for equation (\ref{G(u)k=1}). 
Since we are interested in the solutions for which the relevant
exponent vanishes for $u=-1$, we expand $G(u)$ around the south
pole $u=-1$:
\begin{equation}
\label{ansatza}
G(u) = \sum_{n=0}^{\infty} a_n \left(u+1 \right)^n.
\end{equation}
Assuming that $E \neq g$, we derive the following recurrence
relations. The initial condition is
\begin{equation}
 a_1 = - \tfrac{1}{3} \left(E^2 -\left(g-\tfrac{3}{2}\right)^2 \right) a_0,
\end{equation}
and the recurrence relation is given by
\begin{eqnarray}
\label{recurrence}
\nonumber
a_{n+1} = \frac{ 
\left( \left(E-3g \right) n^2 + \left(3E -2g \right) n -\left( E-g
\right) \left(E^2 - \left(g-\tfrac{3}{2}\right)^2 \right) 
\right)}{\left(E-g\right) \left(2n+3 \right) \left(n+1 \right)} a_n \\
+ \frac{g \left( n^2 - \left(E^2 - E -g^2 +2g +\tfrac{1}{4} \right) \right)} 
{\left(E-g \right) \left(2n+3 \right) \left(n+1 \right)} a_{n-1}.
\end{eqnarray}
Since we have an initial condition we expect to be able to reduce this second
order recurrence relation to a recurrence of first order. Therefore, we make
the ansatz
\begin{equation}
a_n = \prod_{i=0}^{n-1} b_i.
\end{equation}
Substituting this ansatz into the recurrence relation
(\ref{recurrence}) we obtain a non-linear first order recurrence for
the $b_n$. However, this can be solved by setting 
\begin{equation}
   b_n = -\frac{\left(2E-2g+ 1 +2 \left(n+1 \right) \right) 
\left(E^2-E+ 2g - g^2 + \tfrac{1}{4} -
   \left(n+1 \right)^2\right)}{\left(2E-2g+1+2n\right) \left(2n+3
   \right)\left(n+1\right)}. 
\end{equation} 
By construction the series (\ref{ansatza}) has the right behaviour at
the singular point $u=-1$. To 
find the behaviour at $u=1$ we have to discuss the radius of
convergence $R_{{\rm conv.}}$ of the series. The radius 
$R_{{\rm conv.}}$ can be calculated as the following limit: 
\begin{eqnarray}
\nonumber
R_{{\rm conv.}} &=& \lim_{n \to \infty} \left| \frac{a_n}{a_{n+1}} \right|, \\
\nonumber
          &=& \lim_{n \to \infty} \frac{1}{ |b_n| },  \\
          &=& 2. 
\end{eqnarray}
Therefore, the series converges for $u \in [-1,1)$. For $u=1$,
equation (\ref{ansatza}) can be written as
\begin{equation}
\label{G(1)}
 G(1) = \sum_{n=0}^\infty a_n 2^n.
\end{equation}
For large $n$, $b_n$ behaves like ${\frac{1}{2}}$. Therefore, $a_n$
is proportional to ${(\frac{1}{2}})^n$ so that the terms in the sum
(\ref{G(1)}) are of order $1$ and do not vanish. Hence, the series
is divergent, unless it terminates.
But because of the special form of $a_n$ and $b_n$ it is easy to see
when the series terminates. In fact, for $n>0$ we can now write $a_n$
as the following product:
\begin{equation}
a_n = \frac{\left(E + g - \tfrac{3}{2} \right) \left(E+g-\tfrac{n}{2}
\right)}{n!~(2n+1)!!}
\prod_{i=1}^{n} \left(E^2-E+ 2g - g^2 + \tfrac{1}{4} - \left(i+1
\right)^2 \right),
\end{equation}
where $(2n+1)!!$ is the product of all odd numbers less than or equal to
$(2n+1)$.
All $a_n$ vanish for $E = \frac{3}{2} -g$ 
which is the constant solution that we have derived
previously. If the term $(E+g-\frac{n}{2})$ vanishes then the
coefficient for $a_n$ vanishes individually. However, if one of the
terms in the product vanishes, say for $n = N_0$, then not only
$a_{N_0}$ vanishes but also all $a_N$ for $N \ge N_0$, {\it i.e.}
the series terminates.

\begin{figure}[!htb]
\begin{center}
\includegraphics[height=125mm,angle=270]{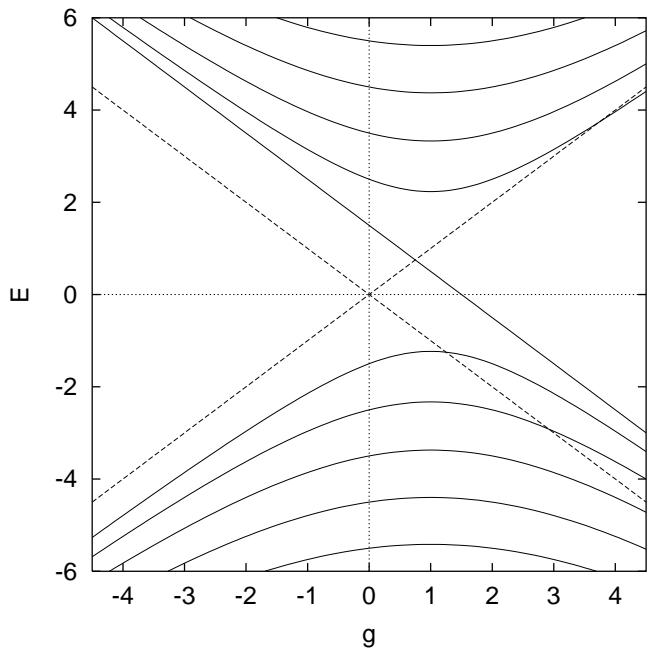}
\caption{The energy levels $E_n$ as functions of the coupling
constant $g$ for $f(\mu) = \mu$. $E = \pm g$ is shown as 
dashed lines.\label{fermion2}}
\end{center}
\end{figure}

From this condition the complete energy spectrum can be calculated.
We obtain the special level $E_0 = \frac{3}{2} - g$ which crosses from
positive to negative energies, and two series of energy levels $E_n^+$
and $E_n^-$ which are given by
\begin{equation}
\label{specf=mu}
E_n^\pm = \tfrac{1}{2} \pm \sqrt{n^2 +2n + \left(g-1 \right)^2}
~~~~~{\rm for}~~~~n = 1,2,\dots.
\end{equation}
The spectrum is displayed in figure \ref{fermion2}.
If we only consider the modes with energy 
$E_n^\pm$ the spectrum is symmetric under
reflection at the axes $g=1$ and $E=\frac{1}{2}$.
However, the mode $0$ with energy $E_0$ is only symmetric under
combined reflections $g \to 2-g$ and $E \to 1-E$, which corresponds to
a $\pi$-rotation around the point $(E=\frac{1}{2},g = 1)$.
This spectrum agrees with our previous calculation where we used the fact that 
$G(u)$ is an integral function.

\section{Numerical Results}
\label{Numerics}

In this section we first describe numerical algorithms to calculate
the spectrum for more general shape functions $f(\mu)$. Then we
discuss the spectrum for the conformal ansatz (\ref{conformal}). We
also discuss the simple ansatz $f(\mu) = B \mu$ which gives rise to
Skyrmions with $B>1$. Finally, we compare our results with the
literature.

\subsection{Numerical Algorithm}
\label{NumericalAlgorithm}

The energy spectrum can be calculated for various shape functions
using a standard relaxation method \cite{NumericalRecipes:1992}. 
The differential equation (\ref{OD1}) is rewritten as a system of
first order differential equations.
\begin{eqnarray}
\nonumber
{y_1}^\prime (u) &=& y_2(u), \\
{y_2}^\prime (u) &=& h\left(y_1(u),y_2(u),E \right).
\end{eqnarray}  
The energy $E$ is considered as an additional dependent variable whose
derivative vanishes, $y_3 (u) \equiv E$, so that the complete system
becomes
\begin{eqnarray}
\label{system}
\nonumber
{y_1}^\prime (u) &=& y_2(u), \\
\nonumber
{y_2}^\prime (u) &=& h \left( y_1(u),y_2(u),y_3(u) \right), \\
{y_3}^\prime (u) &=& 0.
\end{eqnarray} 
For simplicity, let us discuss the numerical algorithm for
$O(4)$-symmetric shape function $f(\mu)= \mu$.
Equation (\ref{G(u)k=1}) has a regular singular point at $u=-1$ which
implies that at this point we cannot choose value and derivative of
the solution independently. 
However, once the value of $G(-1)$ is chosen we can calculate
$G^\prime(u) = \frac{{\rm d}}{{\rm d} u} G(u)$. In fact, inserting $u=-1$ 
into equation (\ref{G(u)k=1}) we obtain
\begin{equation}
G^\prime (-1) = -\tfrac{1}{3}\left(E^2 - g^2 +3 g
-\tfrac{9}{4} \right) G(-1). 
\end{equation}
Similarly we obtain for $u=1$,
\begin{equation}
G^\prime (1) = \tfrac{1}{5}\left(E^2 - g^2 + g -
\tfrac{9}{4} +\frac{3 g}{E+g}\right) G(1).
\end{equation}
The system (\ref{system}) is discretized such that the boundary
conditions at the singular points $u= \pm 1$ are satisfied. 
Starting from an initial configuration the relaxation
algorithm calculates an improved configuration by linearizing the
equation around the solution and minimizing an error function which
indicates how well the discretized equations are satisfied. 

The secret of a successful relaxation method is a good guess for the
starting configuration. Fortunately, for $g=0$ the analytic solution of
(\ref{OD1}) is known for all possible shape functions
$f(\mu)$. Therefore, we start with this configuration at $g=0$ and
increase $g$ by a small amount $\delta g$. The relaxed solution is a
good initial configuration for $2 \delta g$, and so on.

As a test, the spectrum for $f(\mu) = 0$ is calculated using the
relaxation method. This reproduces the analytic result.
For $f(\mu) = \mu$ the relaxation method converges well as long as
$-\frac{E}{g}$ is not inside the interval $[-1,1]$. However, if the
regular singular point $-\frac{E}{g}$ is inside the interval, the
relaxation algorithm no longer converges and has to be modified.
This is one of the main reasons for including the lines $E = \pm g$ in the 
figures.

In this case a specially chosen shooting method proves more suitable. It 
deals with the additional singularities in $[-1,1]$ by simply avoiding 
them. 
As mentioned before, our differential equation can be defined on the
complex plane. 
Due to the boundary condition the solution $G(u)$ is an analytic
function. The original solution is recovered by restricting $G(u)$ to be a
real function of a real variable $u \in [-1,1]$. In order to avoid further
singular points on the real line we can set $u = \exp(i \phi)$ where $\phi
\in [0, \pi]$. Then equation (\ref{OD1}) induces a differential equation
for $\phi$, that is the equation is solved on a semi-circular contour. The
resulting differential equation can be written as a system of 5 real first
order differential equations, one of which corresponds to the energy $E$.
Now, the spectrum can be obtained by a standard shooting method,
\cite{NumericalRecipes:1992}. The boundary conditions for the shooting
method are that at the singular points $u = \pm 1$ the solution is 
non-singular and furthermore that the equation is real at $u = -1$. Since
the equation has real coefficients, the fact that $G(u)$ is 
real for $u=-1$ implies that $G(u)$ is real for $u \in [-1,1]$.

A successful shooting method needs a good initial guess. For $g = 0$ we
can rely on our explicit solution and impose $G(-1) = 1$ and
\begin{eqnarray}
\label{initial1a}
G(1) &=&
\frac{\Gamma(\tfrac{3}{2})\Gamma(-\tfrac{3}{2})}{\Gamma(e)\Gamma(-e)},
\\
\label{initial1b}
&=& - \tfrac{2}{3} e \sin(\pi e)
\end{eqnarray}
where equation (\ref{initial1a}) follows from equation (\ref{hypid}) and
equation (\ref{initial1b}) follows
from the reflection relation for the gamma function (see (6.1.17) in
\cite{Abramowitz:1972}). As in the relaxation method the initial guess for
the coupling constant $g + \delta g$ is the value of the solution for
coupling constant $g$.

It is also possible to reconstruct the solution $G(u)$ for $u \in [-1,1]$
using the solution $G(\exp(i \phi))$ for $\phi \in [0,\pi]$. Cauchy's
formula states that if $f(z)$ is an analytic function and $C$ is a contour
enclosing the point $a$ then 
\begin{equation}
f^{(n)}(a) = \frac{n!}{2 \pi i} \oint\limits_{C}
\frac{f(z){\rm d} z}{(z-u)^{n+1}},
\end{equation}
see for example \cite[page 90]{Whittaker:1927}.
In our case $G(u)$ is a real function on the interval
$[-1,1]$, so 
\begin{equation}
G(u) = \sum\limits_{k=0}^\infty a_k u^k,
\end{equation}
where the coefficients $a_k$ are real. It follows that $G(u)$ can be
written as
\begin{equation}
G(u) = \frac{1}{\pi} \Re \left( 
\int\limits_{0}^{\pi} \frac{G({\rm e}^{i \phi})
{\rm e}^{i \phi} {\rm d} \phi}
{{\rm e}^{i \phi} - u}
\right), 
\end{equation}
and a similar formula can be obtained for its derivative $G^\prime(u)$.
For $E^2 > g^2$ both the relaxation method and the 
shooting method give the same result.

\subsection{Generalization to more complicated Skyrmions}
\label{Generalization}

With the conformal shape function (\ref{conformal}), equation (\ref{OD1}) 
gives rise to a rather lengthy differential equation with five regular singular 
points, see \cite{Krusch:2001} for more details. 
Assuming that the singular points do not coalesce we obtain the following
exponents: 
\begin{eqnarray}
\nonumber
\rho_{1,2}^{(1)}  &=& \left(0,-\tfrac{3}{2} \right),  \\
\nonumber
\rho_{1,2}^{(-1)} &=& \left(0,-\tfrac{1}{2} \right),  \\
\nonumber
\rho_{1,2}^{(\frac{E a - g b}{E b - g a})} &=& \left(2,0 \right),    \\
\nonumber
\rho_{1,2}^{(\frac{a}{b})} &=& \left( 0,0 \right),    \\
\rho_{1,2}^{(\infty)} &=&
\left(\tfrac{3}{2} \pm \sqrt{E^2 - g^2}\right),
\end{eqnarray}
where $a=k^2+1$ and $b=k^2-1$ and $k>1$.
The relation (\ref{erelation}) is again satisfied. Moreover, the
exponents at infinity are the same as for $f(\mu) = 0$.
In the limit $k \to 1$ the exponents at infinity change
discontinuously (see (\ref{exk=1})).
The exponents change if one or more singular points coalesce.
However, in general, there are five singular points which all have
different characteristics.

The regular singular points at 
$u = \pm 1$ provide the boundary conditions for the regular solution,
and are not very sensitive to the exact form of the shape
function. They imply that the solution of this equation locally still is an 
analytic function of $u$.
One might expect that the regular singular point at infinity could be used 
to calculate the degree of the polynomial as in the previous two sections. 
However, for general $k$, the solutions are not regular on the entire
complex plane, so that the exponents at infinity do not contain any
information about the spectrum. This is 
because there is a further singular point at $u= \frac{b}{a}$ with 
exponents $(0,0)$. Generically, the solution diverges 
logarithmically at this singular point such that $u = \frac{b}{a}$ is a 
branch point, see equation (\ref{log}).
The existence of polynomial solutions might be  related to the
spherical $O(4)$ symmetry of the problem, which is broken for $k \neq 1$.
Note that the singular point at $\frac{a}{b} =
\frac{k^2 + 1}{k^2 -1}$ is independent of $E$, and it can
never be inside the interval $(-1,1)$ because $(k^2 + 1)^2 > (k^2 -1)^2$ 
for $k \in (0,\infty)$.
Finally, the singular point at $\frac{E a - g b}{E b - g a}$ depends on
the energy $E$. Moreover, for certain values of $E$, $k$ and $g$ this
point can be inside the interval $(-1,1)$. This singular point could
force a potential solution to diverge, such that
it would no longer be well-defined. Therefore, we are interested
in the local behaviour near $u = \frac{E a - g b}{E b - g a}$. 
After a straightforward but long calculation using
Maple, we obtain $g_2=0$.
Therefore, locally, both solutions are convergent power series, and no
logarithmic divergencies occur.  

\begin{figure}[!htb]
\begin{center}
\includegraphics[height=125mm,angle=270]{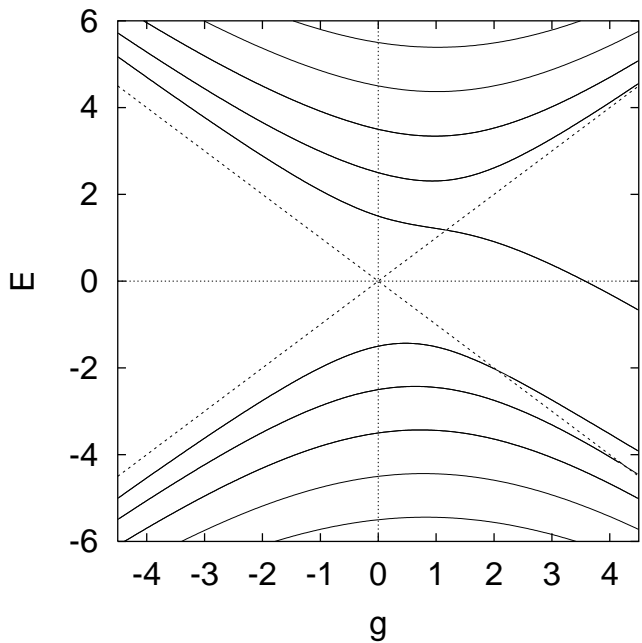}
\caption{The spectrum using (\ref{conformal}) for the shape
function $f(\mu) = 2 \arctan \left(k \tan \frac{\mu}{2} \right)$ where 
$k=5$. $E = \pm g$ is shown as dashed lines.\label{fermion3}}
\end{center}
\end{figure}

In the following we calculate spectra as functions of the coupling
constant $g$ for different values of $k$ numerically. 
For $k=1$ we reproduce the spectrum for $f(\mu) = \mu$ of Sect. 
\ref{HedgehogSkyrmion}.
Also note that $k = \infty$ corresponds to the spectrum for $f(\mu) = \pi$
which agrees with the spectrum for $f(\mu) = 0$ in Sect.
\ref{ConstantSkyrmion}. 
As $k$ increases the
spectrum moves towards the $k=\infty$ spectrum. At the same time, the
energy $E_0$ intersects $E=0$ for increasing values of the coupling
constant $g$, such that for $k \to \infty$ the intersection point also
appears to move to $\infty$. In figure \ref{fermion3} we display the
spectrum for $k=5$.

In the following we discuss the eigenfunction which crosses zero in more 
detail. We restrict out attention to the following Lorentz scalar $Q$ 
which is defined as 
\begin{equation}
Q = \int\limits_{S^3} {\bar \psi} \psi.
\end{equation}
A short calculation shows that ${\bar \psi} \psi$ is a function of $\mu$ 
only. Therefore, we can integrate out the angular coordinates $\theta$ and 
$\phi$ and obtain:
\begin{eqnarray}
\label{Qtilde}
Q  &=& \int\limits_{0}^{\pi} {\tilde Q}(\mu)~ {\rm d} \mu,\\
\label{Qtilde2}
&=& 4 \pi \int\limits_{0}^{\pi} \left(
(1-\cos\mu)~G^2 + 
(1+\cos \mu)~{\tilde G}^2
\right)
\sin^2 \mu~ {\rm d} \mu.
\end{eqnarray}
The density ${\tilde Q}(\mu)$ measures where the fermion field is 
localised.
To obtain equation (\ref{Qtilde2}) we used equations (\ref{0+spinor}) and 
(\ref{FGtilde}).
$G(u)$ and ${\tilde G}(u)$ can be calculated from (\ref{OD1}) and 
(\ref{OD2}), respectively. Recall $u=\cos \mu$.
Expressing $F(u)$ in terms of ${\tilde G}(u)$  avoids 
having to worry about the denominator in (\ref{F}) which is rather 
difficult to handle numerically. 
Note however, that $F(u)$ is non-singular for 
$E+g\cos f = 0$. This follows from the behaviour at the 
corresponding regular singular point of equation (\ref{OD1}).
The function ${\tilde G}(u)$ is related to $G(u)$ via equations 
(\ref{FGtilde}) and (\ref{F}). This implies
\begin{equation}
\label{bc}
{\tilde G}(1) = -\frac{3}{2(E+g)}.
\end{equation}

In the following, we calculate ${\tilde Q}(\mu)$ for $k=1$, $2$ and $5$
and compare it to the baryon density 
\begin{equation}
{\tilde B}(\mu) = \frac{2}{\pi} f^\prime(\mu) \sin^2 f(\mu).
\end{equation}
which measures where the Skyrme field is localized.\footnote{Usually, 
${\tilde B}(\mu)$ is defined with a minus sign, but this is of no 
relevance here.}

The expression for $\psi$ can be calculated explicitly for $k=1$. In this 
case $G(u) = a_0$. Since we are interested in the zero crossing mode we 
have to set $E = \frac{3}{2} - g$ and obtain
\begin{equation}
{\tilde Q}(\mu) = 8 \pi a_0^2 \sin^2 \mu.
\end{equation}
Surprisingly, this is independent of $g$. Also note that in this case 
${\tilde Q}(\mu)$ is proportional to ${\tilde B}(\mu)$. In the following, 
we normalize ${\tilde Q}(\mu)$ so that $Q=1$.

\begin{figure}[!hb]
\begin{center}
\subfigure[${\tilde B}(\mu)$ for $k=2$.]{
\includegraphics[height=55mm,angle=270]{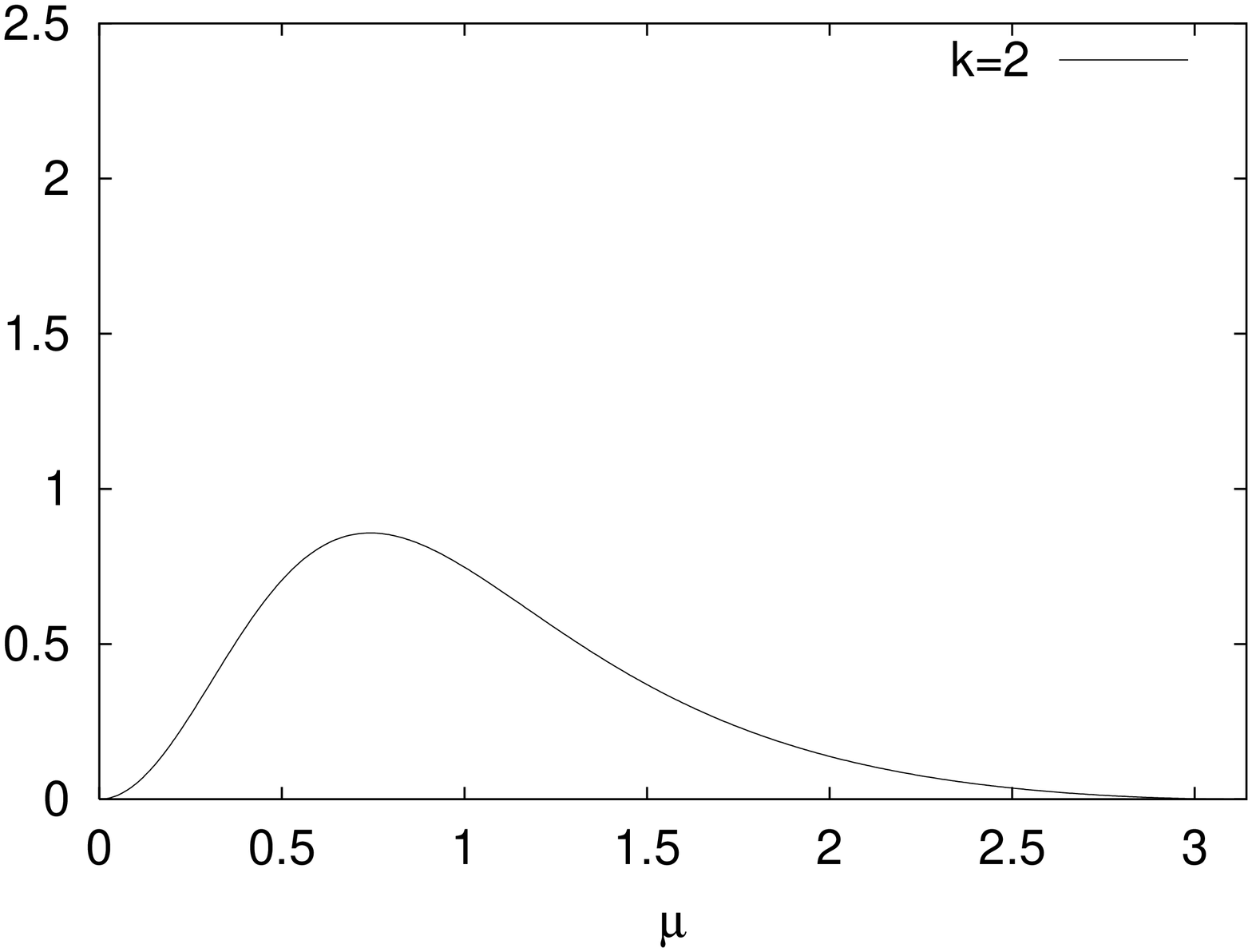}
}
\quad
\subfigure[${\tilde B}(\mu)$ for $k=5$.]{
\includegraphics[height=55mm,angle=270]{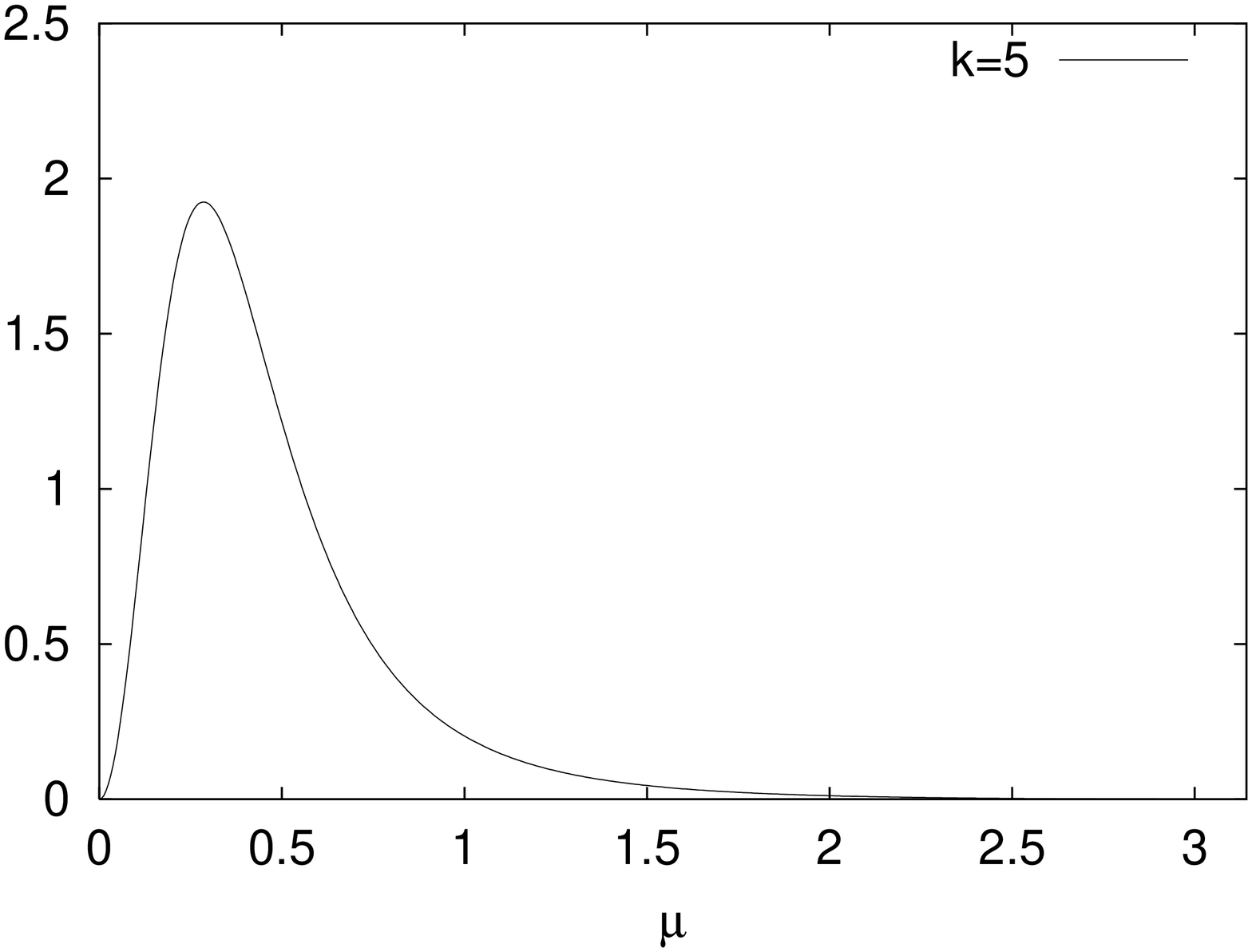}
}
\quad
\subfigure[${\tilde Q}(\mu)$ for $k=2$ for $g=1,2,3,4,5$.]{
\includegraphics[height=55mm,angle=270]{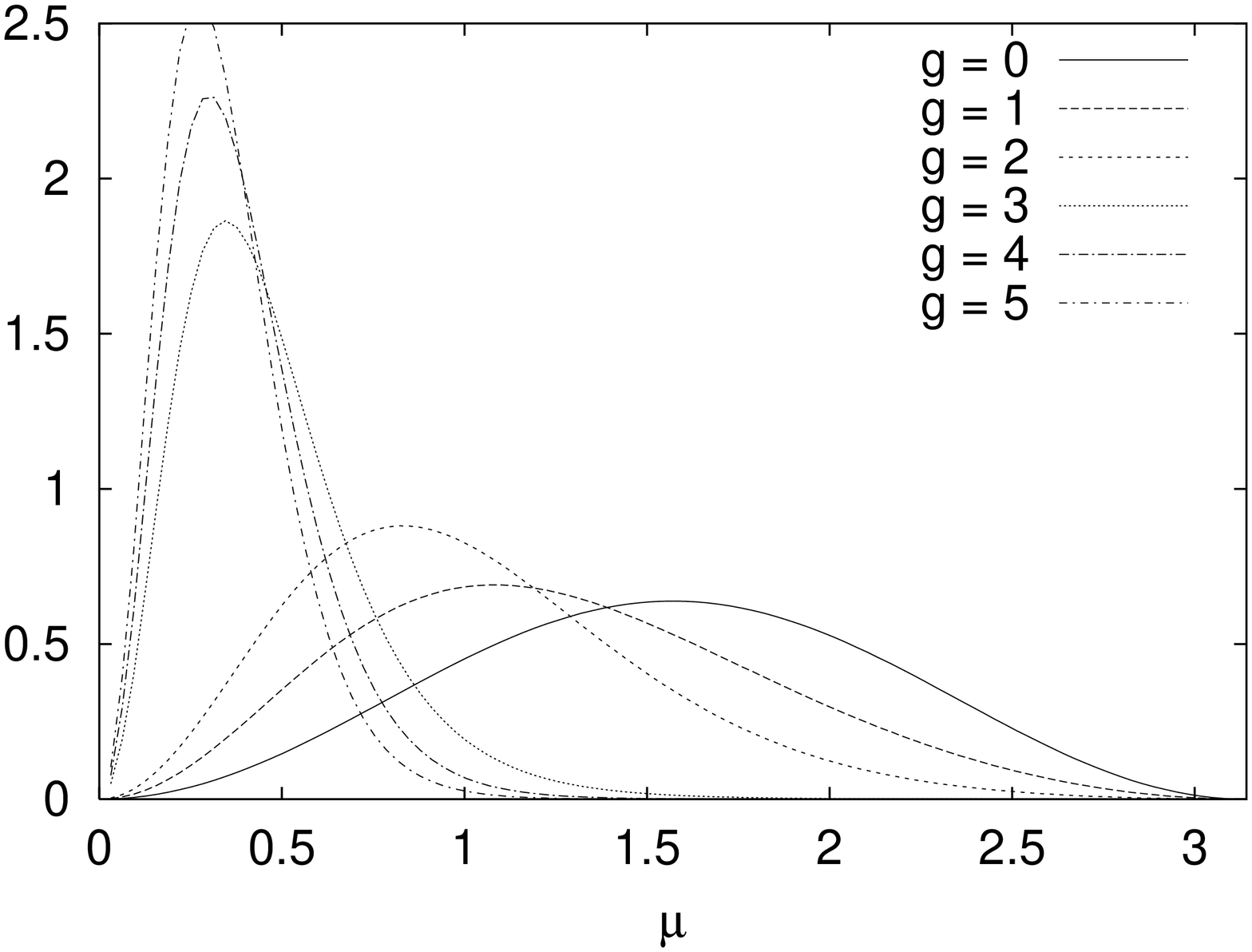}
}
\quad
\subfigure[${\tilde Q}(\mu)$ for $k=5$ for $g=1,2,3,4,5$.]{
\includegraphics[height=55mm,angle=270]{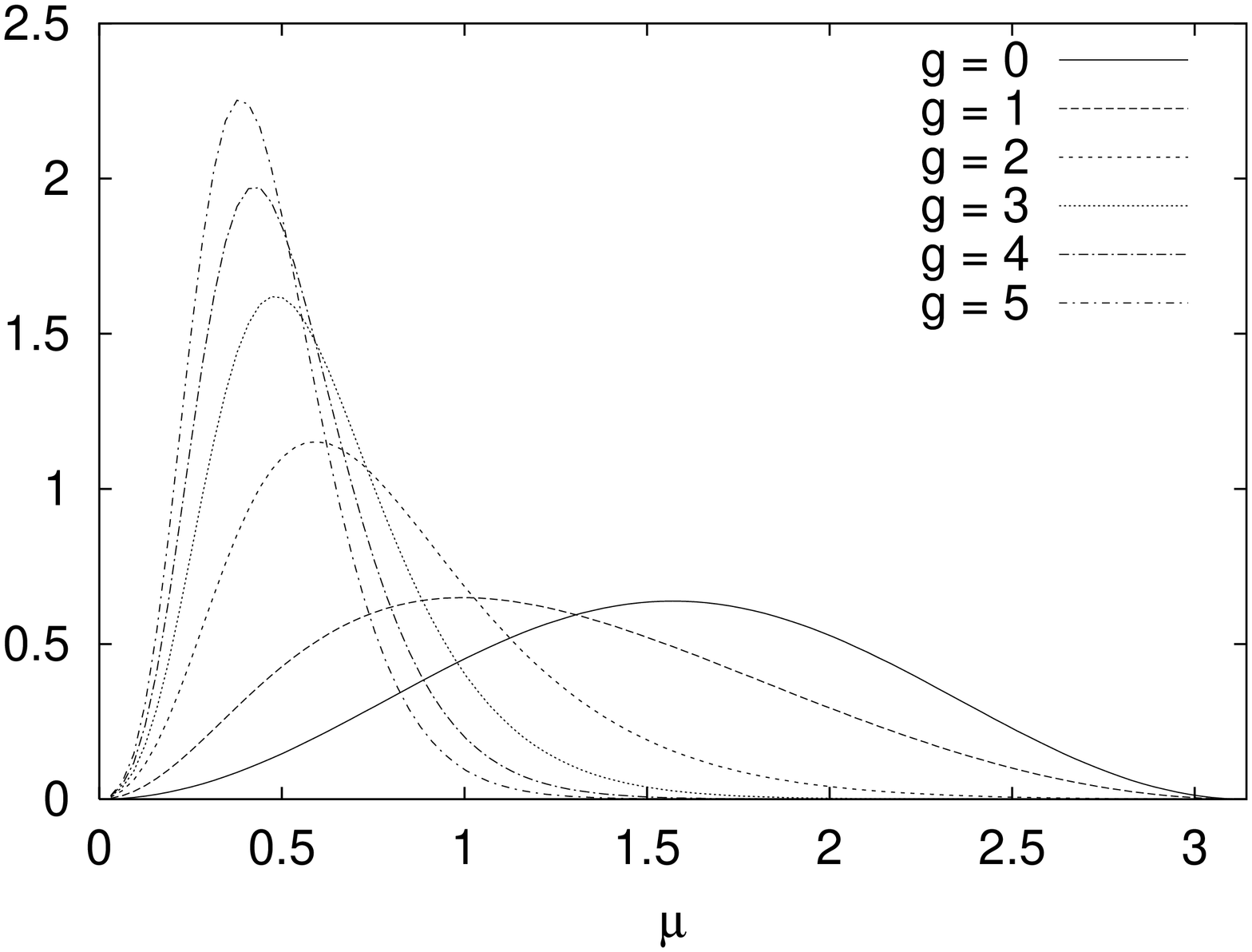}
}
\quad
\subfigure[Comparison of ${\tilde Q}(\mu)$ and ${\tilde B}(\mu)$ for 
$k=2$.]{
\includegraphics[height=55mm,angle=270]{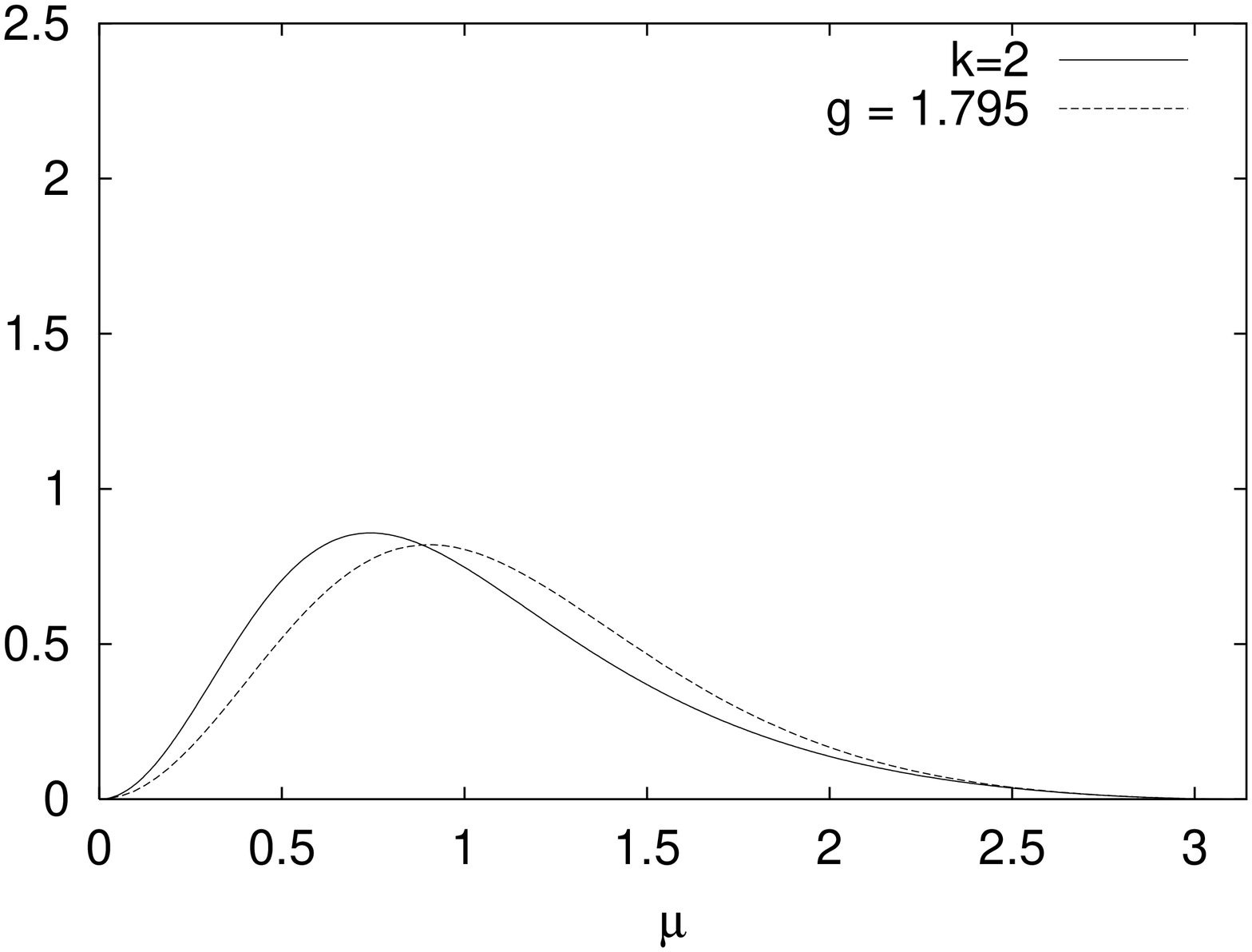}
}
\quad
\subfigure[Comparison of ${\tilde Q}(\mu)$ and ${\tilde B}(\mu)$ for
$k=5$.]{
\includegraphics[height=55mm,angle=270]{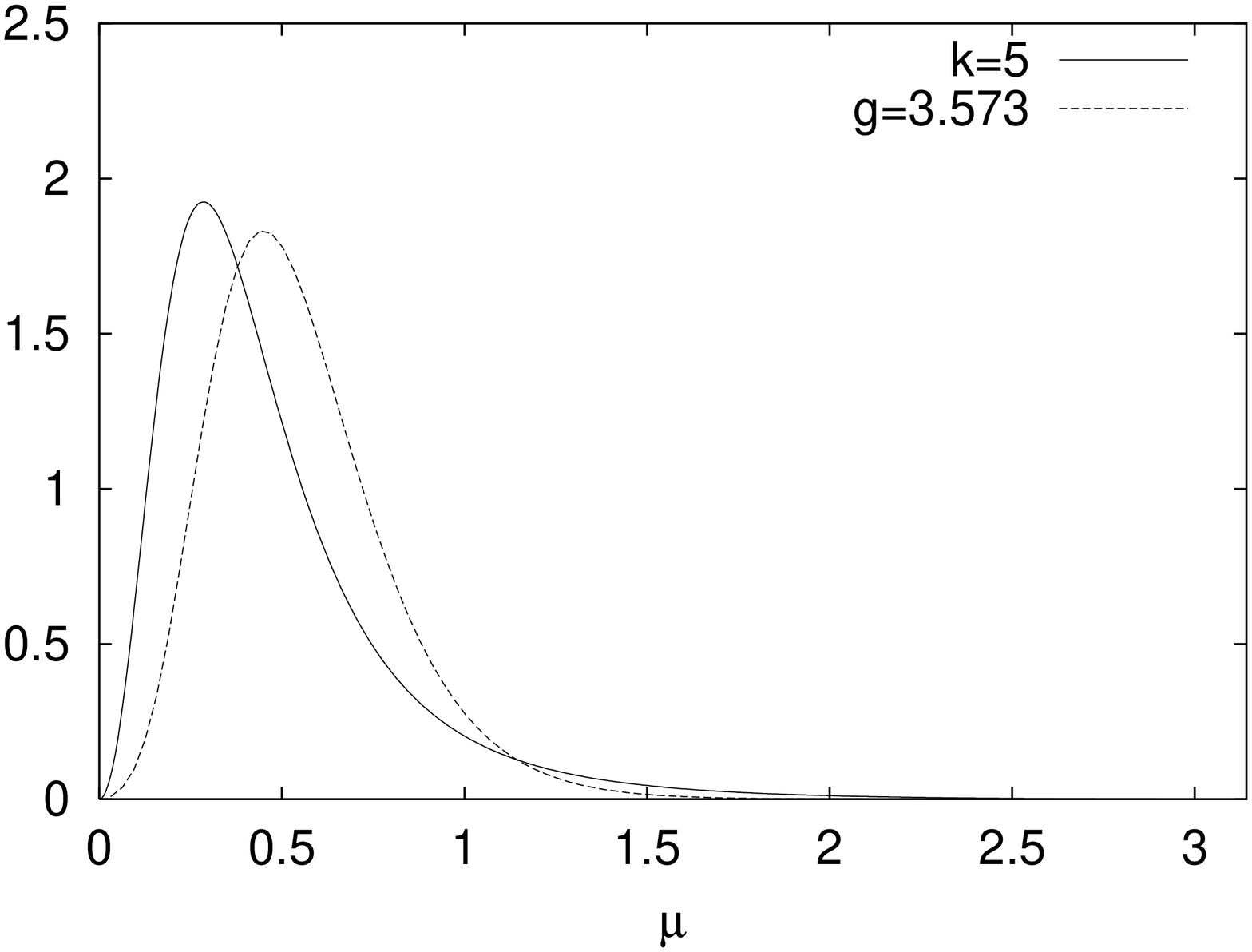}
}
\caption{In this figure we display ${\tilde B}(\mu)$ and 
${\tilde Q}(\mu)$ as a function of $\mu$ for $k=2$ and 
$5$, and for various values of the coupling constant $g$.\label{fermion4}} 
\end{center}
\end{figure}

For $k >1$, the Skyrme configuration is localised near the north pole 
$\mu=0$ (see figure \ref{fermion4}). In figure \ref{fermion4} (a) and (b) 
we show ${\tilde B}(\mu)$ for $k=2$ and $k=5$, respectively. In figure 
\ref{fermion4} (c) and (d) we show ${\tilde Q}(\mu)$ for several values of 
the coupling constant $g$. For $g=0$, ${\tilde Q}(\mu)$ can also be 
calculated 
explicitly and is independent of $k$. For $k>1$ and $g>0$, the fermion 
configuration is also localised near the north pole. For $k=2$, the energy 
$E$ vanishes when $g \approx 1.795$. For $k=5$, the energy $E$ vanishes 
when $g \approx 3.573$. In figure \ref{fermion4} (e) and (f) we display 
${\tilde Q}(\mu)$ for these values of $g$ and compare it to the baryon 
density ${\tilde B}(\mu)$. There is quite a good agreement between these 
two functions. Therefore, the $E=0$ states can be interpreted as a 
fermion-Skyrmion bound states.

The simplest method for obtaining a Skyrmion with baryon number $B>1$
is the ansatz $f(\mu)=B\mu$. For $B>1$ this Skyrme configuration has
very high energy. However, we are mainly interested in the interplay of topology
and the spectrum of the Dirac equation which should be independent of
the precise form of the Skyrmion.

\begin{figure}[!htb]
\begin{center}
\includegraphics[height=125mm,angle=270]{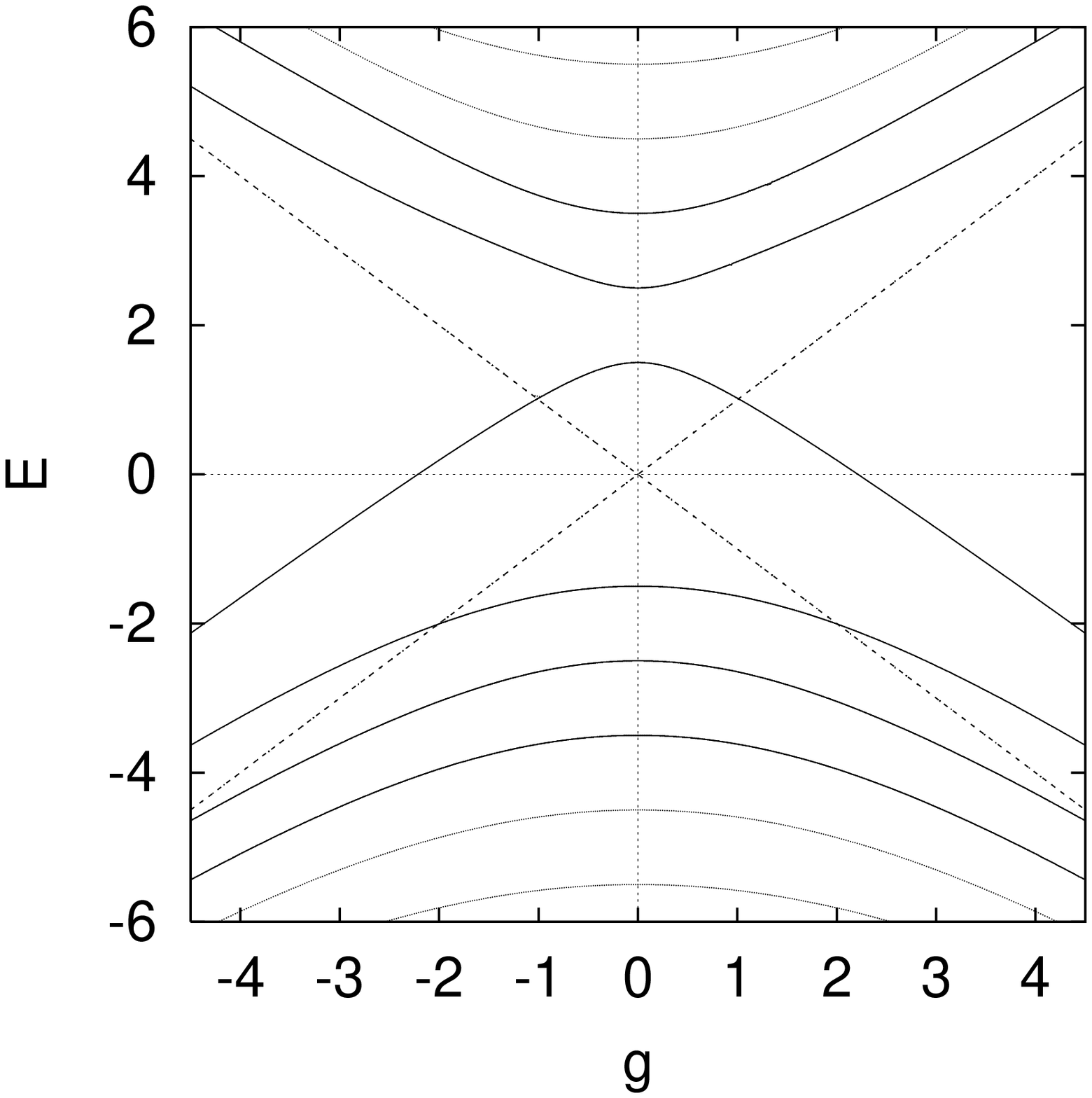}
\caption{The spectrum using $f(\mu) = 2 \mu$. $E = \pm g$ is shown as dashed 
lines. \label{fermion5}}
\end{center}
\end{figure}

For $B=2$ the spectrum is displayed in figure \ref{fermion5}. It 
is symmetric with respect to the $E$-axis. 
Note that the symmetry of the spectrum can be derived from equation (\ref{OD1}) 
and (\ref{OD2}). For even $B$ the function $\cos f$ is an even function of $u$ 
and $\sin f$ is an odd function of $u$.
Under $g \to -g$ and $u \to -u$ equation (\ref{OD1}) transforms into equation 
(\ref{OD2}). However, the equations are just expressed in different 
coordinate systems, and hence have the same energy eigenvalues. Therefore, we 
have shown that the spectrum is symmetric under $g \to -g$ if $B$ is even, 
provided that $f$ is an odd function of $u$.

For $g=0$ all the eigenmodes have the same value,
independent of the background configuration, and parity. As $g$
increases, the modes increasingly feel the influence of the background
configuration, in particular its topology. It appears that the
winding number $B=2$ of the Skyrmion forces $2$ modes to cross zero, one for
$g>0$ and one for $g<0$. In order to follow the spectrum for $E^2 < g^2$ 
the shooting method has to be used.

\begin{figure}[!htb]
\begin{center}
\includegraphics[height=125mm,angle=270]{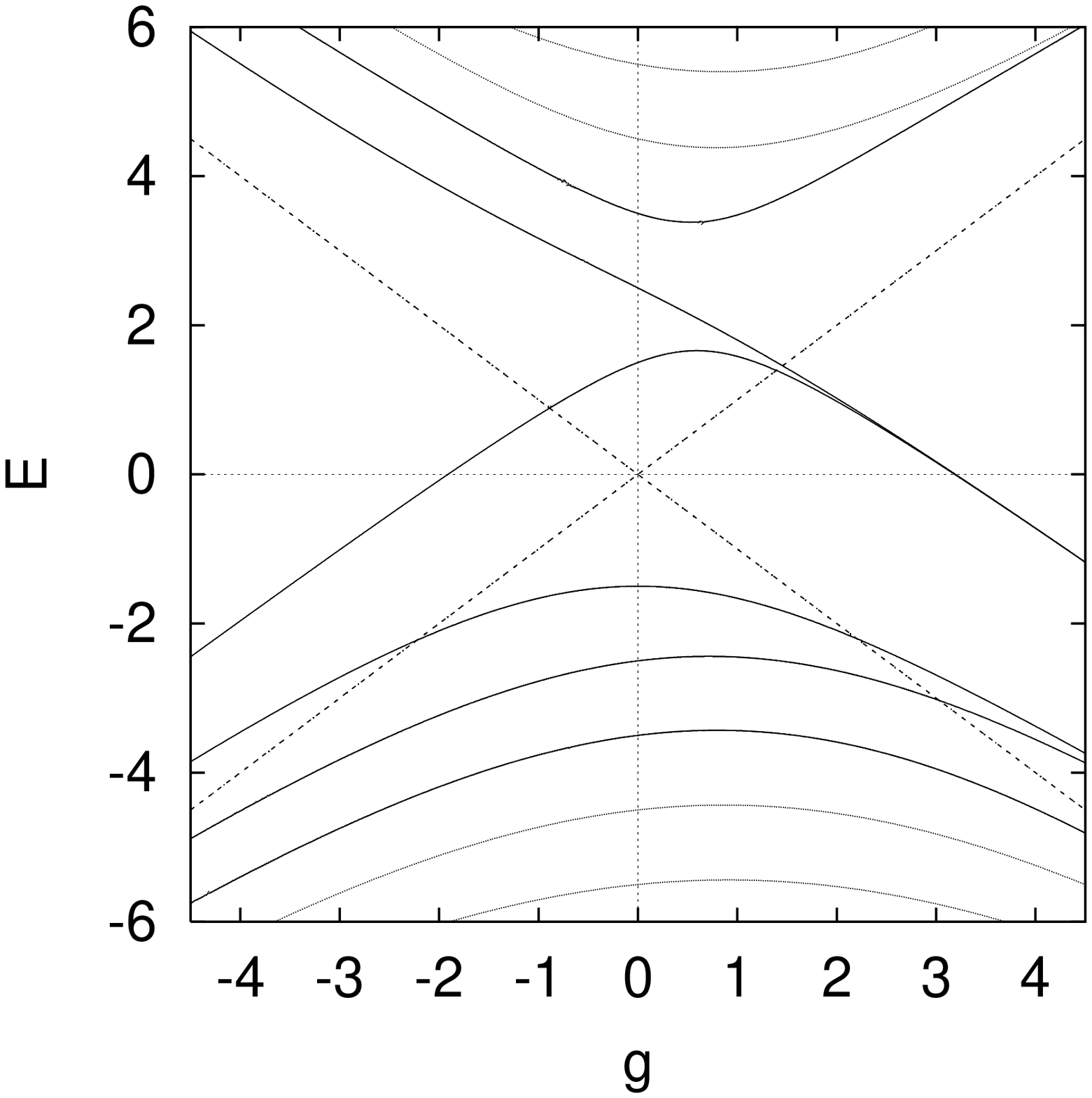}
\caption{The spectrum using
$f(\mu) = 3\mu$. Here we only show the spectrum for $|E| > |g|$, see text. 
$E = \pm g$ is shown as dashed lines.\label{fermion6}}
\end{center}
\end{figure}

This pattern can also be verified for $B=3$, see figure \ref{fermion6}. 
Now there are three modes
which cross zero, two of positive parity ($g>0$) and one of negative
parity $(g<0)$. It seems probable that the number zero-crossing modes
of positive parity differs at most by one from the number of
zero-crossing modes of negative parity. Further calculations for
$B=4$ and $5$ confirm this pattern. For $B=4$ we find two
zero-crossing modes of positive parity and two of negative parity,
whereas for $B=5$ there are three positive parity modes and two
negative parity modes.

This numerical evidence shows that the singular points of equation (\ref{OD1}) 
play an important role. In the following we will assume that $f(\mu)$ is an 
analytic function, and in particular that it can be expanded in a Taylor series 
with respect to the variable $u= \cos \mu$, for $u \in (-1,1)$.
It follows that $\cos f$ and $\sin f$ are also analytic functions of $u$.
Therefore, the only singular points of (\ref{OD1}) inside the interval $(-1,1)$ 
occur for $\cos f = -E/g$. 
If such a singular point is non-degenerate its exponents
are $0$ and $2$. Furthermore, it can be shown be shown by expanding $f$ as a 
power series in $u$ that the coefficient $g_2$ always vanishes. 
In our previous examples $f(\mu)$ was always monotonic. Let $f(\mu)$ be 
monotonic with boundary conditions $f(0)=0$ and $f(\pi) = B \pi$. 
If $E^2 < g^2$ then $\cos f = -E/g$ has $B$ nondegenerate solutions for $u_i 
\in (-1,1)$. 
In this case, these singularities of equation (\ref{OD1}) all have exponents 
$(2,0)$ and the coefficient $g_2$ vanishes, so that the solution of (\ref{OD1}) 
is regular at these points. 
We conjecture that the existence of $B$ regular singular points inside
the interval $(-1,1)$ indicates that there are $B$ modes crossing from one
side of the spectrum to the other. This conjecture is equivalent to 
the statement that for topological charge $B$ there are $B$ modes that cross 
zero.\footnote{The existence of such zero-crossing modes has also been 
discussed recently by Balachandran {\it et al.}  in a more geometric 
language. The authors discovered a relation between these modes and 
Euclidean instantons.}

If we allow for more general Skyrme configurations than the hedgehog,
such a statement becomes even more difficult to prove, because now we
have to  solve partial differential equations. However, both the
topological number and the number of singular points are stable under
small perturbations. Therefore, there seems to be a deep interplay
between the topology of the configuration and the singular points of a
differential operator. This can be compared to monopoles where we have
an index theorem which connects the topology of the monopole
configuration with the index of the Dirac operator.

In the following we compare our results with the literature.
Kahana and Ripka considered the problem of coupling a hedgehog
configuration to fermions in flat space using an approximation
for the Skyrme configuration with a size parameter. Varying this
parameter they computed the spectral flow of the eigenvalues.
They found a $0^+$ zero mode for a special value of the ``size'' of
the hedgehog. 
Since the size is given by minimizing the energy density
for the Skyrmion their result corresponds to varying the
coupling constant $g$.  

In \cite{Kahana:1984be} a linear approximation for the shape
function, $f(r) = \pi(1-\frac{x}{X})$, was used.
Here $x = g \cdot r$ so that the Dirac equation for the fermions is
independent of $g$. However, we defined the Skyrmion as a given
background configuration which minimizes the energy. Therefore,
$X=g \cdot R_{{\rm Sk.}}$ where $R_{{\rm Sk.}}$ is the size of the
Skyrmion. $R_{{\rm Sk.}}$  
can be calculated by minimizing the energy \cite{Adkins:1987kj}: 
$R_{{\rm Sk.}} = 1.67$. 
Kahana and Ripka showed by numerical calculation that there is a zero  
mode for $X=3.2$ which corresponds to $g=1.9$. 
In \cite{Ripka:1997} an exponential function was used to approximate the
shape function: $f(r) = \pi \exp({-\frac{x}{X}})$.
In this case minimizing the energy results in $R_{{\rm Sk.}}=1.7$ and the zero
mode occurs for $X=2$ which corresponds to $g=1.1$.
The fact that there is a mode which crosses zero seems to be
independent of the special ansatz for the shape function. 
This is consistent with our results.

In \cite{Krusch:2000gb} it has been argued that a Skyrmion in flat space 
is reasonably well approximated by the stereographic projection of a Skyrmion on 
a $3$-sphere of optimal radius. The optimal radius for $B=1$ is $L=1$ so we can 
use our analytic result for the mode $E_0$ in Sect. \ref{HedgehogSkyrmion}. 
For $g = \frac{3}{2}$ this is a zero mode, and the value of $g$ is in the same 
order of magnitude as in the literature mentioned above.

\section{Conclusion}

We have derived the Dirac equation for fermions on the $3$-sphere which are
chirally coupled to a Skyrmion. In this paper, we only considered spherically 
symmetric Skyrme configurations and fermionic states with the total
angular momentum  ${G = 0}$. 
We were particularly interested in the shape
functions $f(\mu) = 0$ and $f(\mu) = \mu$ and the conformal ansatz
(\ref{conformal}).  

The shape function $f(\mu)=0$ leads to a hypergeometric equation and can 
therefore be solved explicitly in terms of hypergeometric
functions. To derive the spectrum we have to impose that the solutions
are non-singular at the north and south pole using the correct chart.
We derived the spectrum
\begin{equation}
\nonumber
 E = \pm \sqrt{g^2 + (N+\tfrac{3}{2})^2}~~~~~{\rm for}~~~~~
N = 0,1,2,\dots,
\end{equation}
which agrees with the spectrum calculated by Sen
\cite{Sen:1986dc} using Maurer Cartan forms. Moreover, the
eigenfunctions $G_N(u)$ are given by Jacobi Polynomials.

The shape function $f(\mu) = \mu$ is more complicated. 
It leads to a Fuchsian equation with four regular singular
points, two at the boundary $u = \pm 1$, one at infinity and one
depending on $E$ and $g$. However, the equation could still be solved in
terms of polynomials. We were able to solve the recurrence relation
for the coefficients of the polynomial and derived the corresponding
spectrum: 
\begin{eqnarray}
\nonumber
E_0 &=& \tfrac{3}{2} - g, \\
\nonumber
E_n^\pm &=& \tfrac{1}{2} \pm \sqrt{n^2 +2n + \left(g-1 \right)^2}
~~~~~~{\rm for}~~~~~~ n = 1, 2,\dots
\end{eqnarray}
The eigenfunctions have been evaluated explicitly:
\begin{eqnarray}
\nonumber
G_0(u) &=& a_0, \\
\nonumber
G_n(u) &=& \sum\limits_{k=0}^{n}  
a_k \left(u+1\right)^k,
\end{eqnarray}
where
\begin{equation}
\nonumber
a_k = \frac{\left(E + g - \tfrac{3}{2} \right)
\left(E+g-\tfrac{k}{2} \right)} {k!~(2k+1)!!}
\prod_{i=1}^{k} \left(E^2-E+ 2g - g^2 + \tfrac{1}{4} - \left(i+1
\right)^2 
\right).
\end{equation} 
We also considered more general shape functions numerically. For $B=1$
we discussed the conformal ansatz (\ref{conformal}) and calculated the
spectrum. 
For $k=1$ the spectrum agrees with the analytic solution. In
particular there is a mode $0$ whose energy $E_0$ crosses zero for a
certain value of the coupling $g$. 
The graph of $E_0$ as a function of $g$ changes as the 
parameter $k$ in the ansatz is varied. However, for finite $k>0$, this 
mode always stays present. 
According to our numerical evidence the eigenfunction for $E=0$ can be 
interpreted as a fermion-Skyrmion bound state.

For larger $B$ we considered the ansatz $f(\mu)=B\mu$. This leads to
$B$ modes crossing zero. We conjectured a connection between zero modes
and certain singular points in the equation which in turn are
connected to the winding number of the configuration.

\section*{Acknowledgements}

The author wants to thank N S Manton, H Pfeiffer and Y Shnir for many 
discussions during various stages of this project. 
The author also would like to thank J M Speight for many helpful 
suggestions and continuous support.

\renewcommand{\baselinestretch}{1}
\addcontentsline{toc}{section}{Bibliography}
\begin{small}

\end{small}
\label{lastref}
\end{document}